\pdfoutput=1
\documentclass[smallextended,natbib]{svjour3}\usepackage[]{graphicx}\usepackage[]{color}
\makeatletter
\def\maxwidth{ %
  \ifdim\Gin@nat@width>\linewidth
    \linewidth
  \else
    \Gin@nat@width
  \fi
}
\makeatother

\definecolor{fgcolor}{rgb}{0.345, 0.345, 0.345}

\usepackage{framed}
\makeatletter
 {\par\unskip\endMakeFramed%
 \at@end@of@kframe}
\makeatother

\definecolor{shadecolor}{rgb}{.97, .97, .97}
\definecolor{messagecolor}{rgb}{0, 0, 0}
\definecolor{warningcolor}{rgb}{1, 0, 1}
\definecolor{errorcolor}{rgb}{1, 0, 0}
\newenvironment{knitrout}{}{} 

\usepackage{alltt}  
\smartqed  
\usepackage{graphicx}
\usepackage{array,multirow,graphicx}
\IfFileExists{upquote.sty}{\usepackage{upquote}}{}
\begin{document}

\title{Clicks and Cliques 
}
\subtitle{Exploring the Soul of the Community}

\author{Natalia da Silva \and
        Ignacio Alvarez
}


\institute{Natalia da Silva \at
              Department of Statistics, Iowa State University 
              \email{ndasilva@iastate.edu}
           \and
           Ignacio Alvarez \at
          Department of Statistics, Iowa State University 
           \email{ialvarez@iastate.edu}
}

\date{Received: date / Accepted: date}
\maketitle

\begin{abstract} 
In the  paper we analyze 26 communities across the United States with the objective to understand what attaches people to their community and how this attachment differs among communities. How different are attached people from unattached? What attaches people to their community? How different are the communities? What are key drivers behind emotional attachment? To address these questions, graphical, supervised and unsupervised learning tools were used and information from the Census Bureau and the Knight Foundation were combined. 
Using the same pre-processed variables as \cite{full2010} most likely will drive the results towards the same conclusions than the Knight foundation, so this paper does not use those variables.

\end{abstract}


\section{Introduction}
\label{intro}
Community attachment is an emotional connection to a place that transcends satisfaction, loyalty, and  passion. Residents that are  attached to a community have strong pride in it, a positive outlook on the community's future, and a sense that it is the perfect place for them \citep{full2010}.
The Knight Foundation report presents ten community metric variables in order to explain the key drivers of emotional attachment. They reported that people are more attached to the community if it is more welcoming of social groups (openness) and if people in the community care about each other (social offering).

\begin{figure}
\begin{knitrout}
\definecolor{shadecolor}{rgb}{0.969, 0.969, 0.969}\color{fgcolor}

{\centering \includegraphics[width=0.48\linewidth]{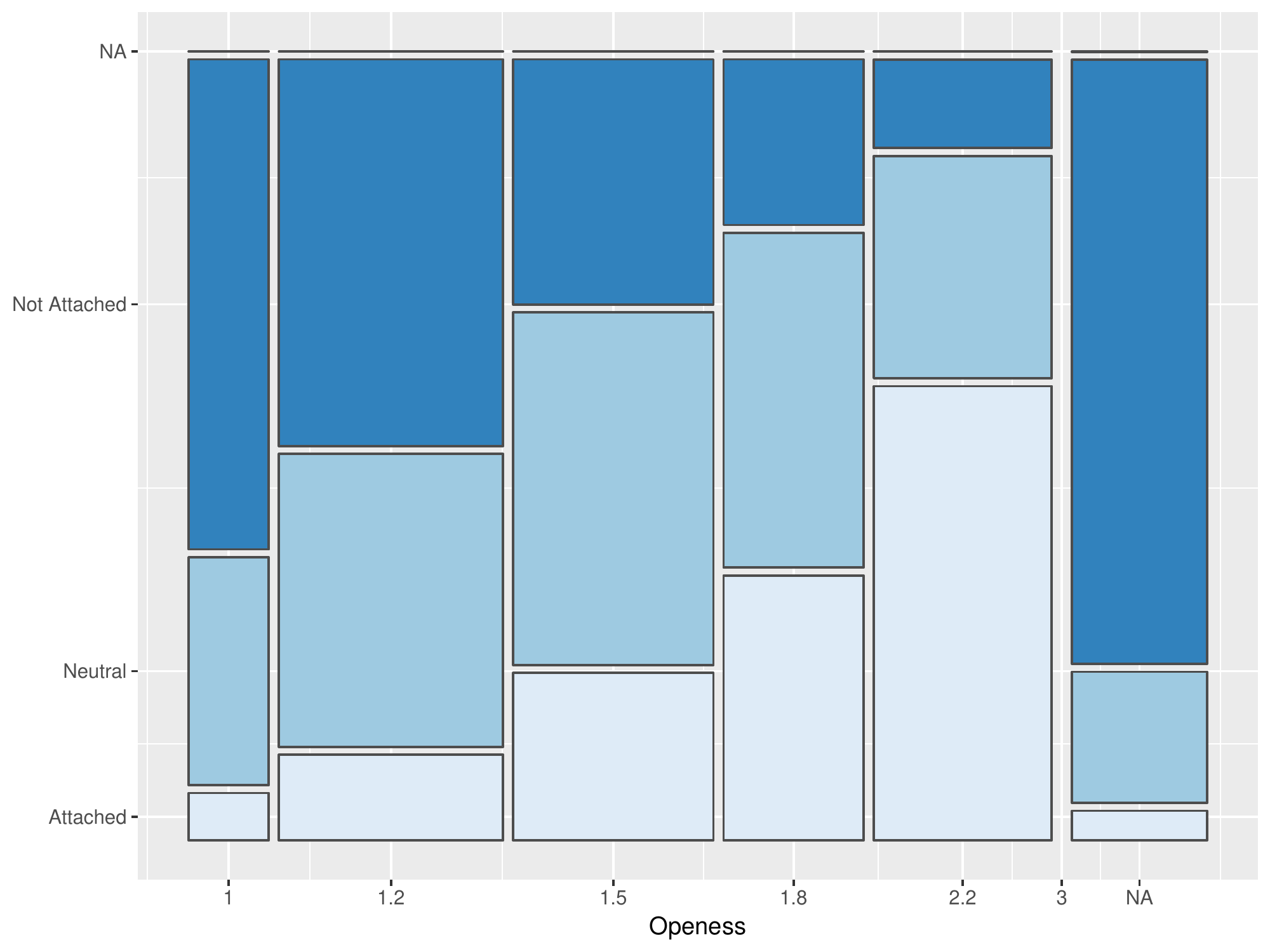} 
\includegraphics[width=0.48\linewidth]{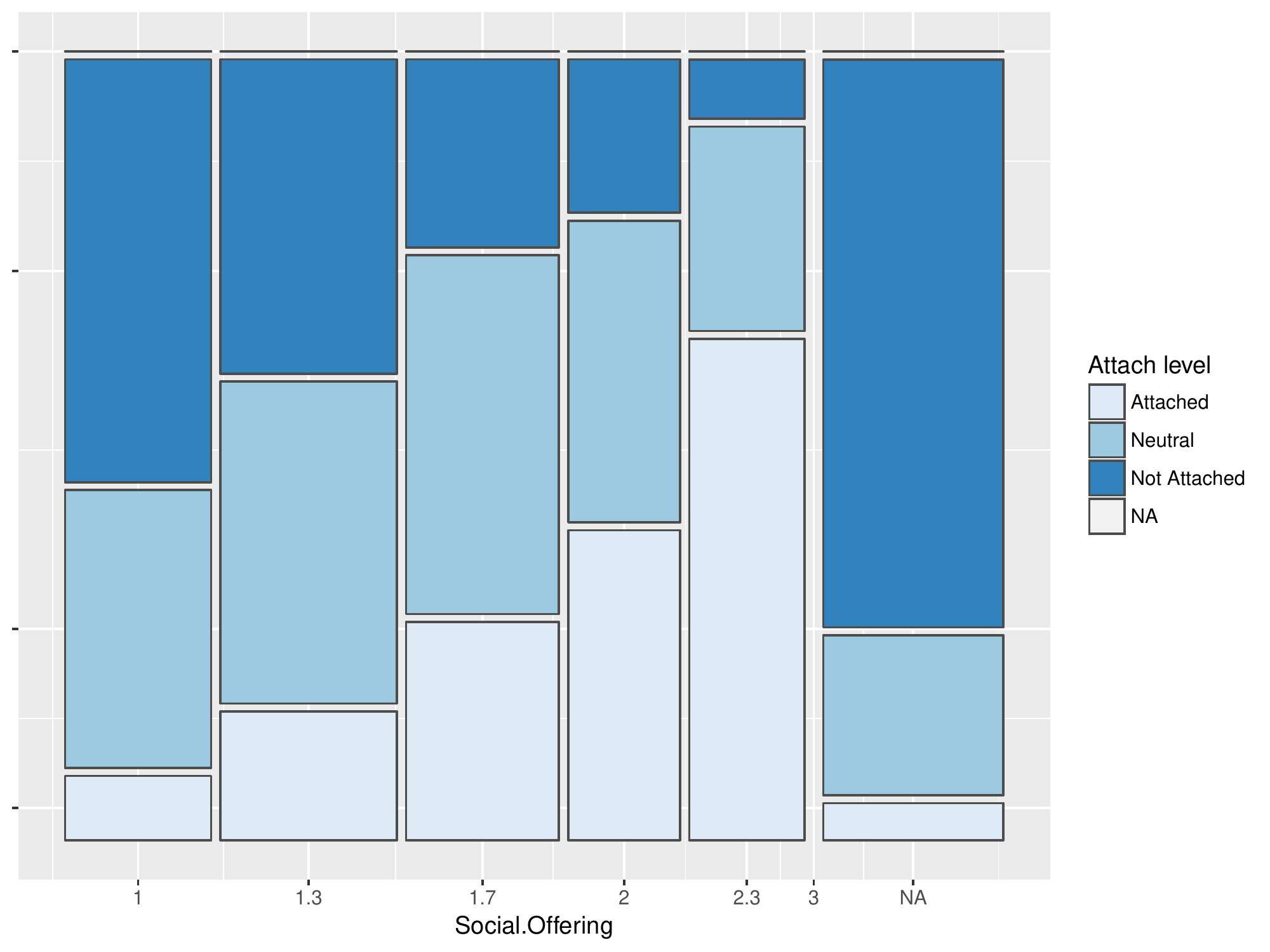} 

}

\end{knitrout}
  \caption{Product plots showing attachment distribution, in the left panel within levels of openness and in right panel within levels of social offering. Openness and Social Offering are binned to make them discrete using quantiles. This plot shows a strong positive association for both openness and social offering and the emotional attachment.\label{pp1}}
\vspace{-.4cm}
\end{figure}

This finding is illustrated in Figure \ref{pp1} with the product plot framework proposed by \cite{wickham2011product} (see Section \ref{methods} for a more detailed description). Attachment is plotted vertically against openness (left) and social offerings (right). The proportion of attachment group conditional to the level of openness (social offering) is represented in the plot. Each vertical bar represents a quantile of openness (social offering), and is divided vertically according to the attachment group. We can see in Figure \ref{pp1} that the proportion of attached people is larger when the openness index is high and also when social offering index is high. According to the Knight Foundation research, this pattern is common for all communities in the study \citep{full2010}.

The challenge for this paper is to identify key drivers of emotional attachment other than those reported by the Knight Foundation. Using the same pre-processed variables as \cite{full2010} it is  likely we will obtain  the  same  conclusions as the Knight Foundation, so we do not use those variables.

We report a deeper exploration of the survey data by looking for new patterns in order to understand what attaches people to their community and how the attachment differs among communities.

The main questions that lead the analysis are:

\begin{itemize} \vspace{-.4cm}
\item What are the key drivers behind community attachment?
\item How different are the communities? Determining if all 26 communities have similar patterns of attachment is crucial to building policy recommendations.
\item How different are attached people from not attached people?
Identifying groups of people and characterizing them in relation to levels of attachment may reveal important aspects of emotional attachment.
\end{itemize}

These questions are addressed at an individual level and at the community level. The attachment variable is measured at an individual level and is analyzed at that level. However, it may be that some community characteristics influence the attachment of the people living in the community, but the variables obtained from the \cite{acs2011} are only available at community level. These two levels of analysis may lead to different perspectives on attachment.

The paper is organized as follows. Section 2 describes the data sources and statistical methods. Section \ref{descri} presents a description of the main characteristics of the communities based on supplementary data obtained from the Census Bureau. Section \ref{explo} explores the relationship between attachment and other variables. Key drivers of attachment are presented in Section \ref{driver} based on the Knight Foundation survey data.

\section{Methods and sources of information}
\label{methods}

Two sources of information are used, data provided by the Knight Foundation and the other from the American Community Survey data from the Census Bureau. This section describes the sources and initial processing of the data, and the main statistical tools used for the analysis.

\subsection{Data description}
The main data set used in this paper was provided by the Knight Foundation, consisting of collected data from 47821 people over three years (2008, 2009 and 2010) in 26 communities across the United States. The data is available at http://streaming.stat.iastate.edu/dataexpo/2013/ as part of the 2013 Data Expo competition (see \cite{soul} for additional details).

Additionally, community level information from the \cite{acs2011} and seven tables of the 2007-2011 American Community Survey 5-Year estimates were obtained to provide a socio-demographic description of each community.

Table \ref{des.var} shows the description of the variable groups from the two sources of information. There are a total of 232 variables in the Knight Foundation data. The variables can be divided into  four main subsets: 1) \emph{original} variables correspond to the survey questions to explain attachment pattern; 2)\emph{demographic} variables, also survey questions, have the objective to characterize demographic aspects of the respondent; 3) \emph{recoded} variables consist of a re-codification of the original ones; and 4) \emph{metric} variables were constructed by the Knight Foundation researchers to explain the attachment process.

On the other hand, each table from the \cite{acs2011} has a different topic (see Table \ref{des.var}). The information corresponds to community totals for different socio-demographic variables, age group, sex, race, education level, average income, employment, etc. Different proportions, rates or indices were computed from these variables, Table \ref{des.var} presents each table description and which new variables were computed from this information.

\begin{table}[h]
\centering
\caption{Sources of information and variables used in the analysis \label{des.var}}
\begin{tabular}{|c|l|p{10cm}|}\hline
 & \textbf{Variable Group}&\textbf{Description}\\ \hline
 \multicolumn{1}{ |c| }{\multirow{4}{*}[-12pt]{\rotatebox[origin=c]{90}{Knight survey data}}}
 & Original variables & There are 118 original variables, most of them consist of ordinal variables with a scale from 1 to 5. All these variables ask for the opinion about some aspect of community life.\\ \cline{2-3}
& Demographic Variables &There are 30 demographic variables, these variables have information about, age, race, gender, etc.\\ \cline{2-3}
 & Recoded Variables&Some of questionnaire variables has been recoded in a scale from 1 to 3. \\ \cline{2-3}
 & Constructed variables & Using recoded variables, some summary variables have been made. These new variables are the base for the Gallup analysis of this data and represent the different sociological dimensions to be study. Some of these variables measure, economy level, social offerings, openness, safety, etc\\ \hline
 \multicolumn{1}{ |c| }{\multirow{7}{*}[-30pt]{\rotatebox[origin=c]{90}{Census Bureau data}}}
& Age-Sex &  Total of people on each age group by gender. Proportions of female and the average age for each gender were computed \\ \cline{2-3}
 & Education & Total of people between 25-64 years old, divided by education level attained and working status. Unemployment rate and the proportion of people within each education level were computed. \\ \cline{2-3}
 & Income &People in each level of income. We computed average income, income per-capita and Gini coefficient\\ \cline{2-3}
 & Owner &Information relative to the total of people living in houses which are renters or are owners, and also how long they are living in that place. We computed average years that people have lived in the same house separately for renters and owners.\\ \cline{2-3}
 & Race &People totals in each race, mainly white, black and Other.\\ \cline{2-3}
 & Workers & Information relative to the household size and the workers per household.\\ \cline{2-3}
 & Year of entry &Total of people which has born outside the USA, and how long they lived in the community.\\ \hline
\end{tabular}
\end{table}

The Knight survey data contains missing information. The proportion of missing values was computed for each observation (by cases) and for each variable (by variables). Table \ref{tab1} shows the quantiles of the proportion of missing values computed by case and computed by variable. Focusing on the proportions of missing values by case, there is 19\% of missing record for the most complete response. This may be explained by the fact that 3 years of data was combined although some variables only appeared in 1 or 2 years. On the other hand, focusing on the proportion of missing values by variables, most of the variables show a reasonable proportion of missing values (the median is 12\%) while there are a few variables with really high proportions of missing values.

\begin{table}[ht]
\centering
\caption{Quantiles of the proportion of missing values by case and variables.} 
\label{tab1}
\begin{tabular}{lccccc}
  \hline
 & Min & Q25 & Median & Q75 & Max \\ 
  \hline
by Variable & 0.00 & 0.10 & 0.12 & 0.68 & 1.00 \\ 
  by Case & 0.19 & 0.25 & 0.33 & 0.37 & 0.91 \\ 
   \hline
\end{tabular}
\end{table}

In order to obtain a data set for the statistical analysis part of the information is not considered. First, only variables within the original and demographic group are used.
Secondly, the set of variables with less than 25\% of missing cases are selected. Among the selected variables, only complete cases are used. The final data set created with this procedure produce 30000 complete observations and 65 variables.

\subsection{Statistical methods}

Supervised and unsupervised learning tools were used in this paper to answer the main questions presented in Section \ref{intro}. To present the results of statistical learning methods, different visualization devices were used to perform an exploratory analysis. In this section the main tools for graphical and statistical analysis are described.

\subsubsection{Product plots}

\cite{wickham2011product} proposed a new framework for visualizing categorical data, where the area is proportional to the count (or proportion) of interest. The framework name, \emph{product plots}, comes from the relation of two types of products, the product of the conditional and marginal distribution to produce the joint distribution, and the product of height and width to produce an area. Then, it is possible to visualize \emph{simultaneously} the marginal probability of a variable and conditional probability with another variable of interest.

Product plot is a very flexible framework, more than 20 visualizations previously described can be included in it \citep{wickham2011product}, like bar charts, spine plots \citep{hummel1996linked} and mosaic plots \citep{hartigan1981mosaics,friendly1994mosaic}. This framework is designed for study relations among categorical variables. However, it is easily adapted when the variables are continuous simply by binning variables.

Considering again Figure \ref{pp1} for describing the relationship among attachment and openness. Each vertical bar consists of a conditional distribution of attachment for one quantile of openness. The width of the bars represent the marginal proportions for openness and each rectangular area represents the joint proportion of the cell.

The main variable of interest in this paper is the attachment level. Product plots are used to study the relationship among attachment and other variables of interest, to characterize communities and cluster of individuals in term of attachment level. In addition, other kinds of visualizations such as scatter plots, parallel plots and tile plots are used.

\subsubsection{Clustering}

Cluster analysis is an unsupervised technique and two ways to find clusters are used across this paper. Hierarchical agglomerative cluster is used to find clusters at the community level. This method starts with each point being its own cluster and they are recursively merged. Then, the ''merging history'' forms a hierarchical tree diagram called a dendrogram, which is a tool used to define the number of clusters (see Sec. 12.3.1 of \cite{izenman2008modern} for more details).

Also, K-means clustering was used to find clusters at the individual level. The K-means algorithm, proposed by \cite{macqueen1967some}, starts with all observations assigned to $K$ initial clusters and cluster means are computed. Then in an iterative fashion observations are reassigned to minimize the within sums of squares of each cluster until a criterion of convergence has been met (see Sec. 12.4.1 of \cite{izenman2008modern} for more details).

In Section \ref{descri} clusters of communities were found using information from \cite{acs2011}. The objective is to look for demographic similarities among the 26 communities. Visual inspection of the dendrogram was used in determining the number of clusters (dendrogram is not shown). In Section \ref{explo}, clusters of individuals with Knight survey data were found. The goal is to see how different attached people are from not attached people. K-means algorithm is used to find clusters of individuals, the number of clusters is determined computing the within sums of squares for consecutive numbers of clusters and deciding the number of cluster where there is no extra reduction in this quantity. Finally, in Section \ref{driver}, an importance index to determine attachment level is computed for each variable and groups of communities are found one more time using the variable importance index.

\subsubsection{Random Forest}
Random forest is a supervised learning technique proposed by \cite{breiman2001random}, based on tree predictors with extra randomness added in its construction. A tree predictor is a classifier which recursively partition the feature space looking for regions where the response variable is homogeneous (pure nodes). In order to build a forest, bootstrap samples from the data are selected and for each sample a tree classifier is grown. However, the constructed tree in each bootstrap sample only uses a random selection of variables to find the next feature partition. Then two sources of randomness are used, random selection of cases in the bootstrap sample and random selection of variables. Then the forest classifier is an ensemble of randomized trees, the forest prediction is determined with majority vote across trees (see Section 14.4 of \cite{izenman2008modern} for more details).

Random forest classifiers are used to classify each individual into its attachment level. The goal is to identify what attaches people to their communities and differences among communities.
For each community, a random forest classifier is fitted with attachment as the response. The goal is not to predict attachment level for one person, but to identify the important variables to predict attachment in one community. By doing this, key drivers of attachment are found. Among the randomized tree classifiers that form the forest, each variable is used many times for splitting the data set. Averaging the decreases in node impurity index over all trees is the Gini measure of variable importance.

\section{General description by communities \label{descri}}
Based on information from the \cite{acs2011}, socio-demographic characteristics of the communities in the study are explored graphically. Knowing the community characteristics can help us to understand the attachment process.

\begin{figure}[hbpt]
\centering
\begin{knitrout}
\definecolor{shadecolor}{rgb}{0.969, 0.969, 0.969}\color{fgcolor}

{\centering \includegraphics[width=\maxwidth]{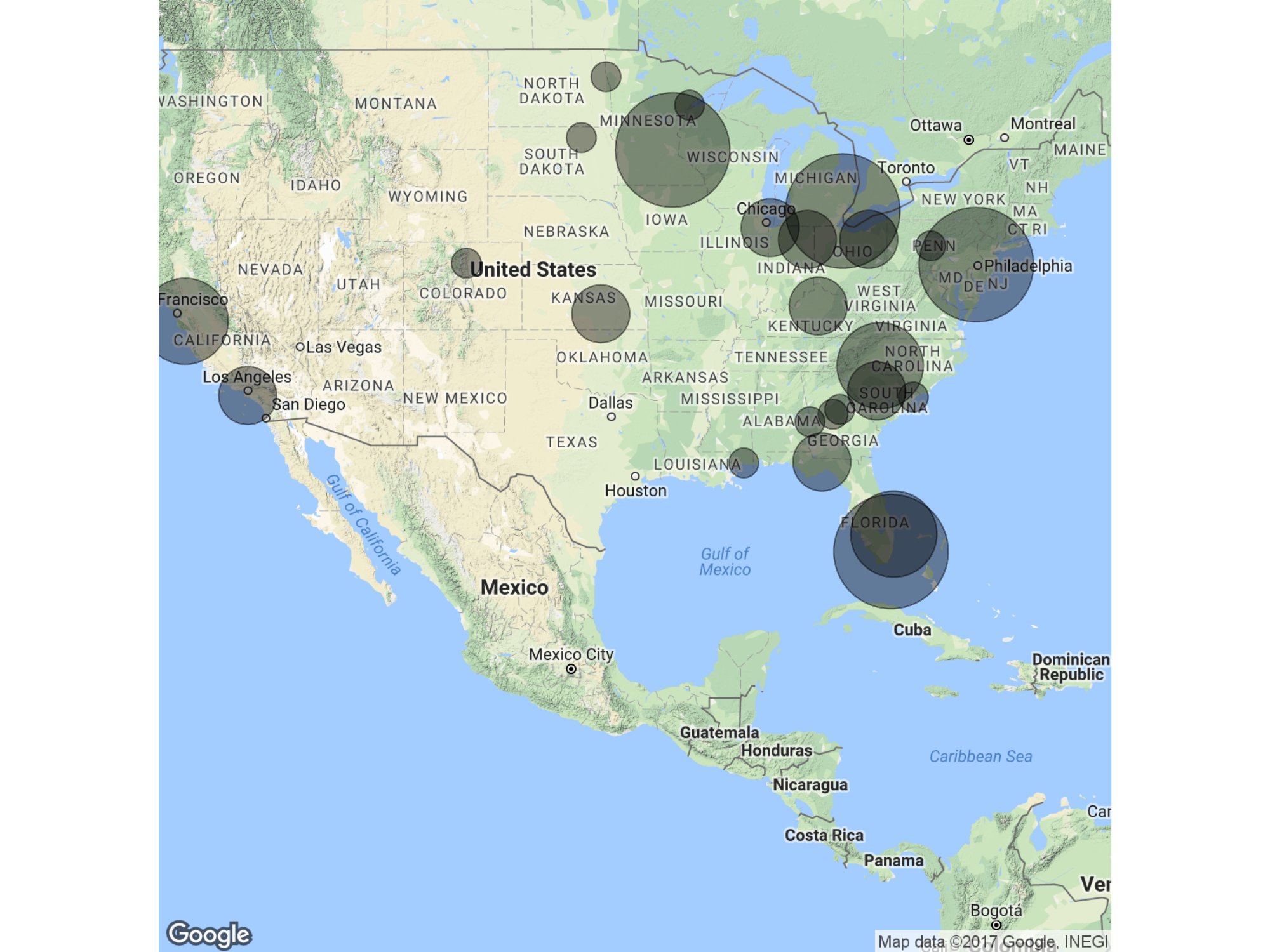} 

}

\end{knitrout}
  \caption{Location of the 26 communities analyzed, size of dot represent population. Only two communities on the west coast. \label{map}}
\end{figure}

Figure \ref{map} shows the communities location, most are from the East Coast of the United States. Size of the dot represents population. Detroit and Philadelphia are the biggest communities while Aberdeen and Milledgeville are the smallest.

\begin{figure}[hbpt]
\begin{knitrout}
\definecolor{shadecolor}{rgb}{0.969, 0.969, 0.969}\color{fgcolor}

{\centering \includegraphics[width=\maxwidth]{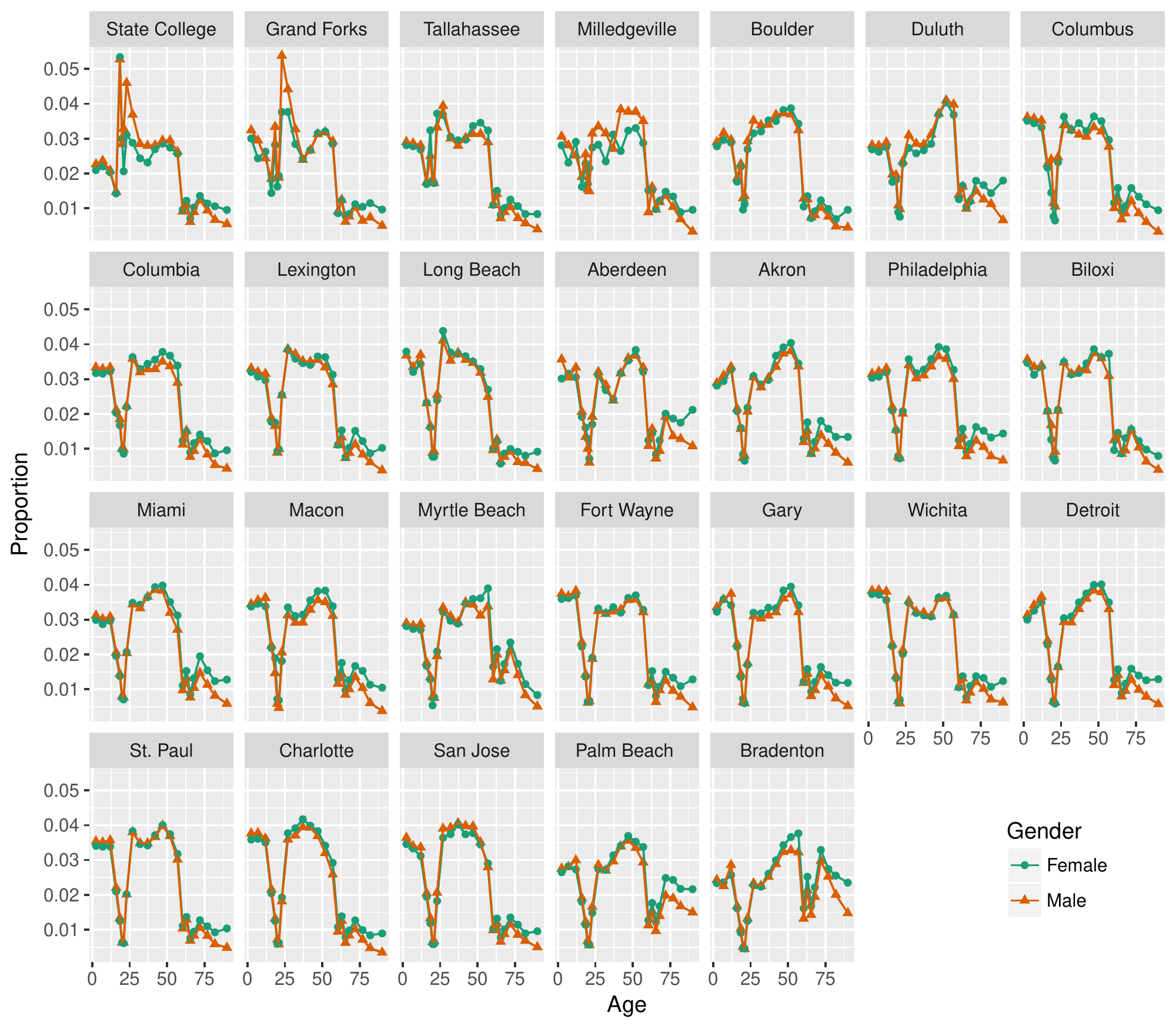} 

}

\end{knitrout}
\caption{Proportion of people by age category, horizontal axis is the midpoint age for each group. The line color and point shape represents the gender. Each facet is a community, sorted from top to bottom according to proportion of people between 18 and 21 years old. Similar age distribution across gender in all communities. Only State College and Grand Folks shows a peak on 18-21 years old. \label{age}}
\end{figure}

The \cite{acs2011} presents a population distribution among age groups and gender. There are 23 categories for age, most of them grouping 5 years together, and in each age category the midpoint is computed as a representative age for plotting. Figure \ref{age} shows the proportion of males and females by age category across all the communities.  The general profiles are very similar for all communities. In most communities, the proportion of people between 18 and 21 is around 5\%. However, there are five communities where the proportion of people between 18 and 21 is larger, Grand Forks (13\%), Milledgeville (12\%), Tallahassee (13\%), Boulder (9\%), and State College (22\%) of the total population. Communities with a larger proportion of people over 55 years old are Palm Beach and Bradenton.

\begin{figure}
\begin{knitrout}
\definecolor{shadecolor}{rgb}{0.969, 0.969, 0.969}\color{fgcolor}

{\centering \includegraphics[width=\maxwidth]{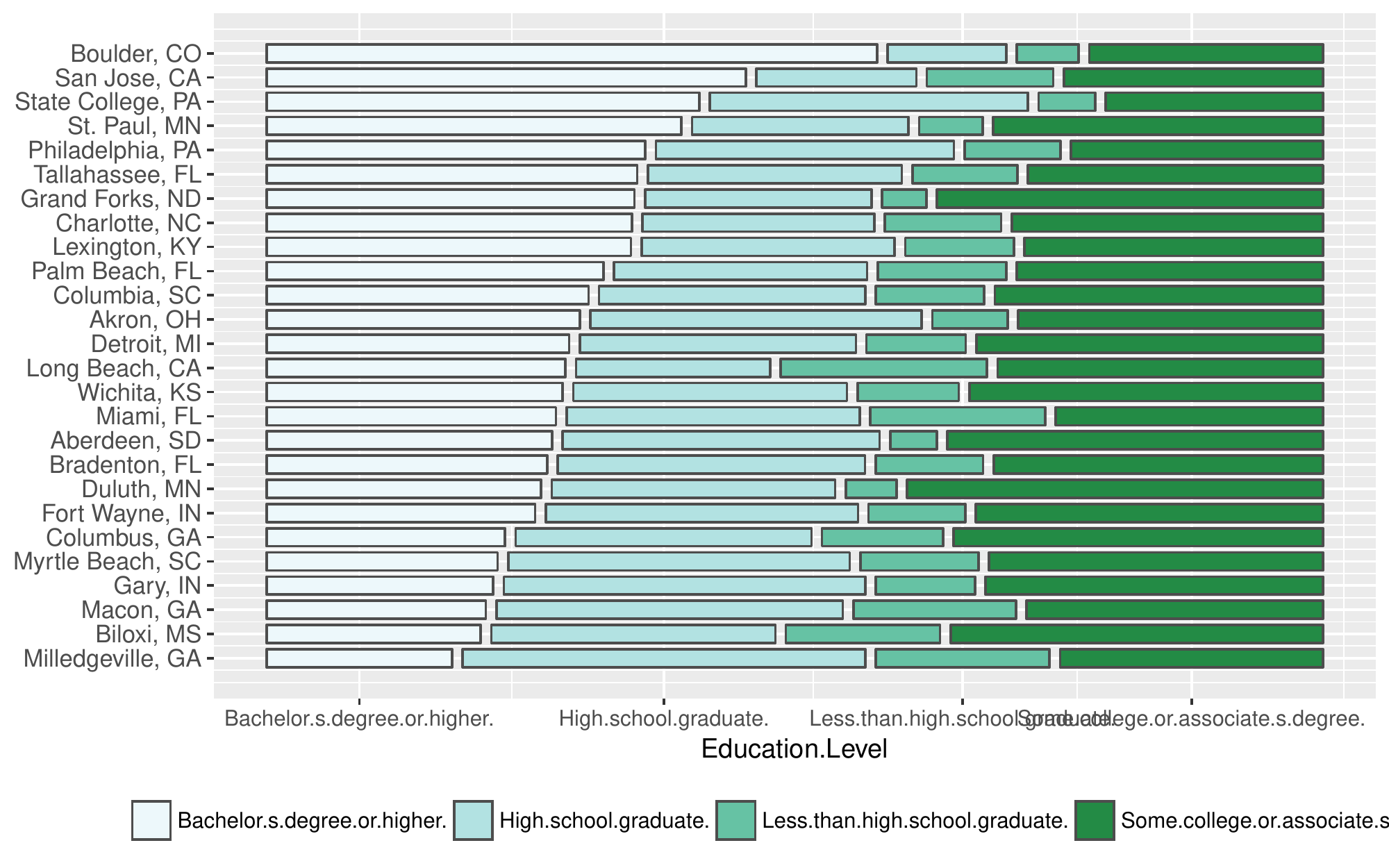} 

}

\end{knitrout}
\caption{Product plot of education level by community, the width of each box represent the proportion of people with that education level within a community.  Communities are sorted from top to bottom by proportion of people with college degree. Big range of variation on education level among the communities. \label{edu}}
\end{figure}

Figure \ref{edu} shows the proportion of people in each education level across communities, sorted from top to bottom by proportion of people with a college degree. Since the data from the \cite{acs2011} is already aggregated at the community level, this product plot only reports proportion of education within each community (not joint proportions), which is why all horizontal bars have the same height.
We can observe that Boulder, San Jose, and State College are the most educated communities while Milledgeville, Biloxi, and Macon are the less educated communities.
\begin{figure}
\begin{knitrout}
\definecolor{shadecolor}{rgb}{0.969, 0.969, 0.969}\color{fgcolor}

{\centering \includegraphics[width=0.49\linewidth]{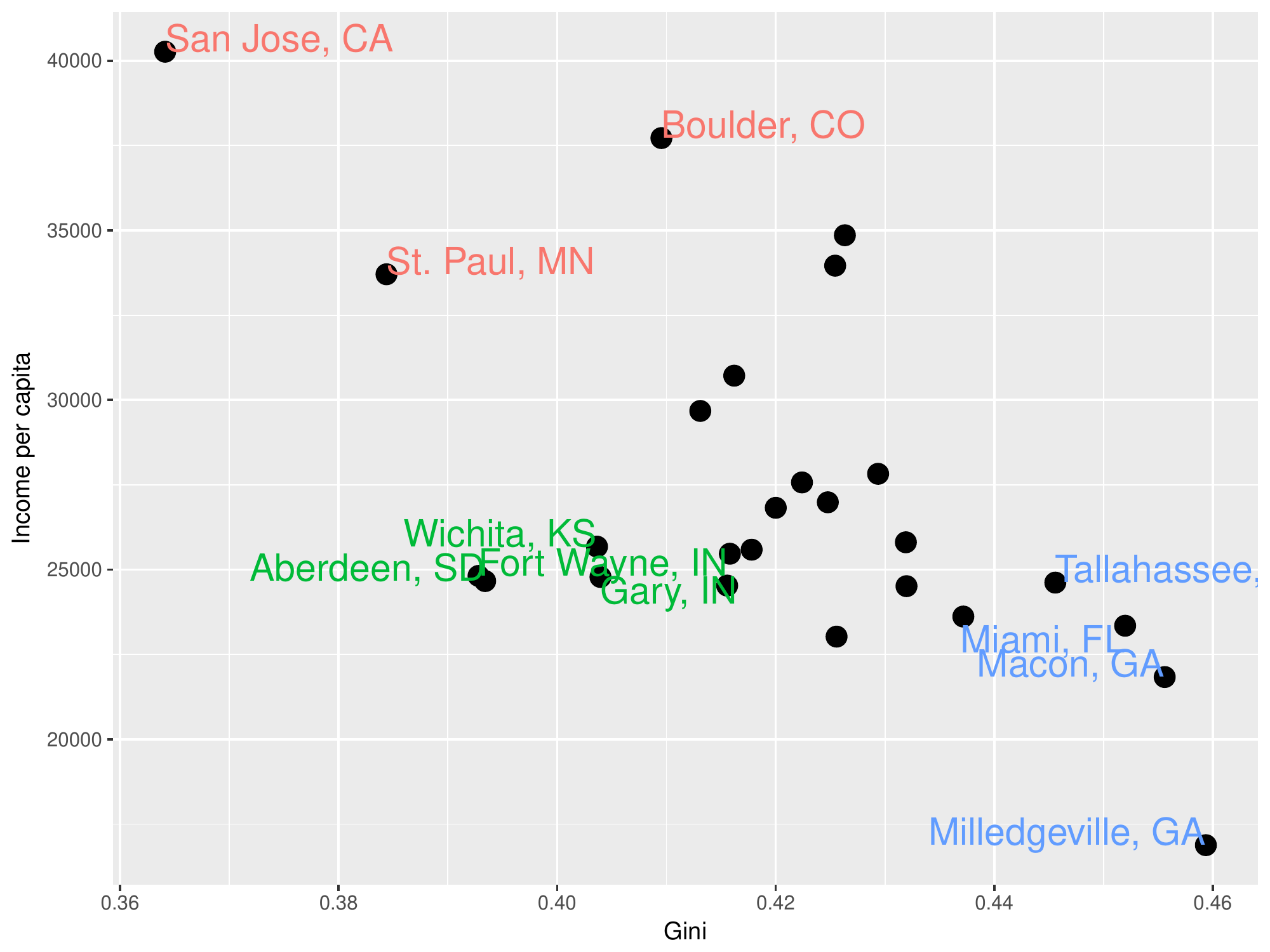} 
\includegraphics[width=0.49\linewidth]{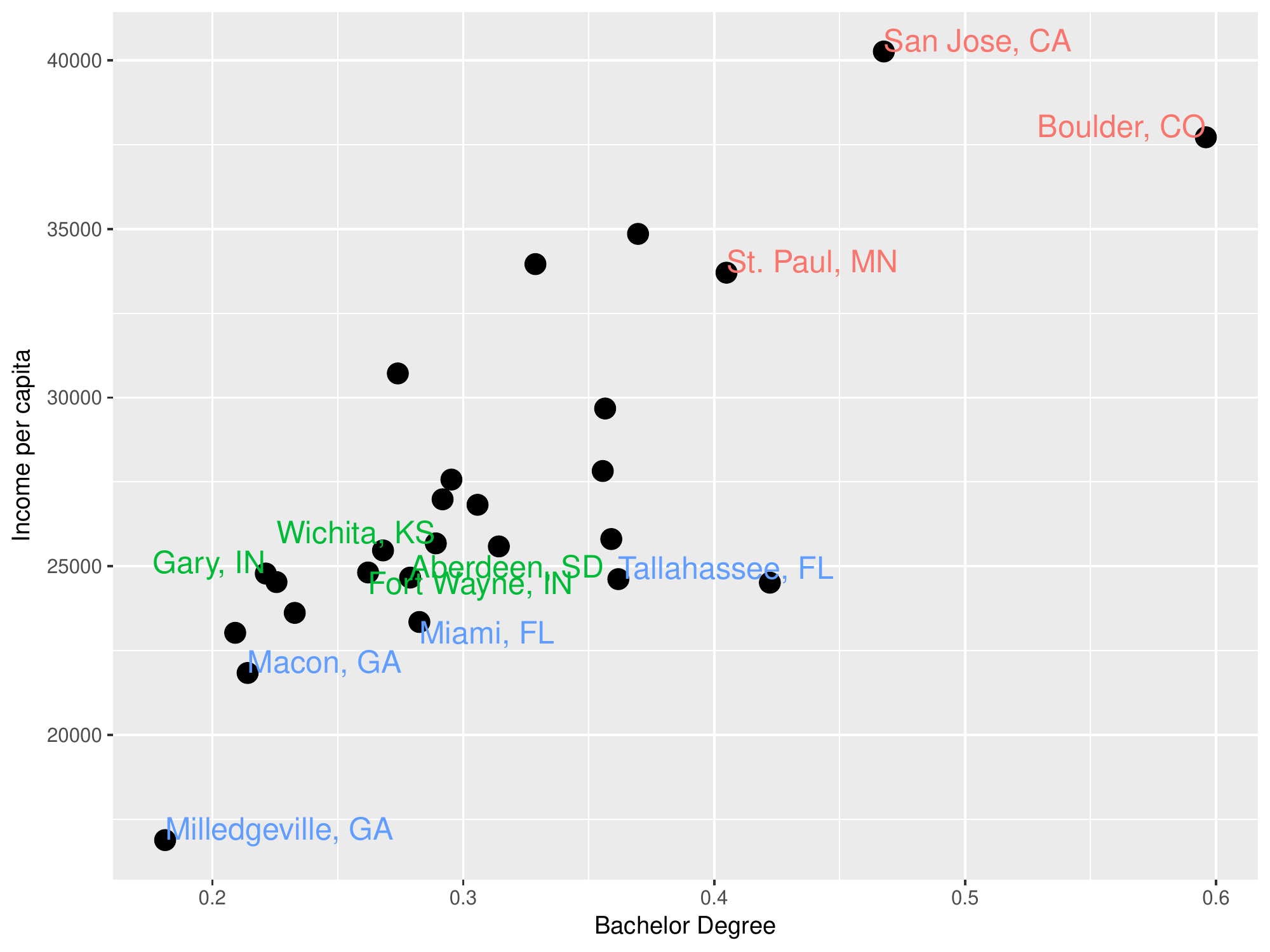} 

}

\end{knitrout}
\caption{Scatter plot of income per capita by Gini index (left) and scatter-plot of income per capita by proportion of people with bachelor degree (right). Colors identify some groups based on income and Gini. \label{ecolink}}
\end{figure}

There exist several measures of income inequality but Gini is the most commonly used in applied studies \citep{atkinson2000handbook}. For this reason the income distribution for each community is summarized using the Gini index. Gini index takes values between 0 and 1, where 0 represents a situation in which every person has the same income level (complete equality) and 1 implies a total income concentration (complete inequality).

Figure \ref{ecolink} shows scatter plots comparing the relationship between income per capita and distribution of income (left), and the proportion of people with a bachelor degree (right).
Different colors identify different groups based on income and Gini. The colored communities are the same in both plots. We can observe that poor communities have more unequal income distribution while in communities where the income per capita is bigger the income distribution is better than in the poor group.

\begin{figure}[hbpt]
\begin{knitrout}
\definecolor{shadecolor}{rgb}{0.969, 0.969, 0.969}\color{fgcolor}

{\centering \includegraphics[width=\maxwidth]{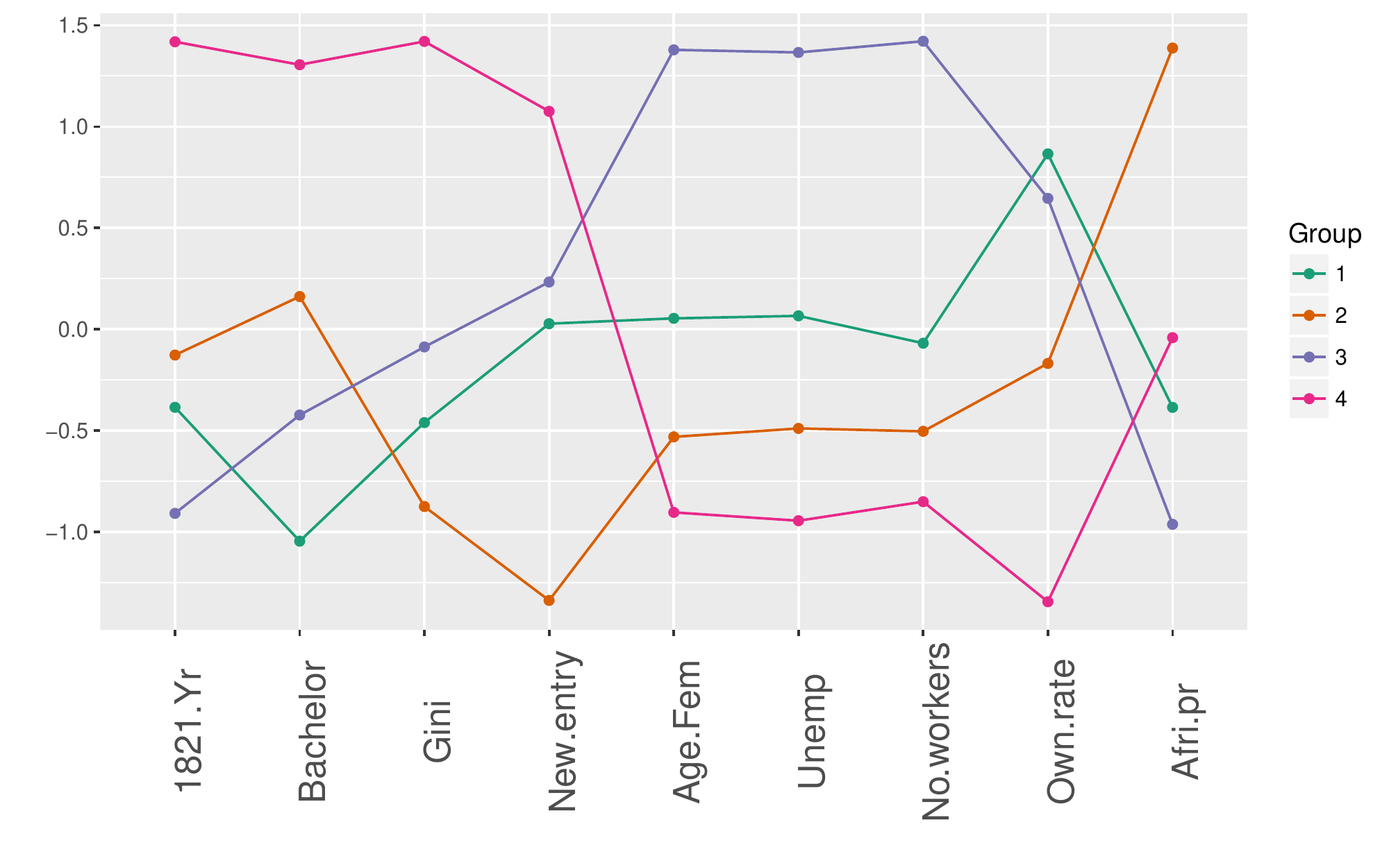} 

}

\end{knitrout}
\caption{Parallel coordinate plot for community cluster means \label{cluster.cen}.  }
\end{figure}

As a final step in the socio-demographic characterization of the communities, cluster analysis on community characteristics was performed. For this analysis nine variables from the Census Bureau data are used: proportion of people between 18 and 21 years old (1821.yr), proportion of people with a bachelor degree (bachelor), the Gini inequality index (Gini), proportion of people who moved to the community after the year 2000 (new.entry), the average age of the female population (age.fem), the unemployment rate (unemp), no workers rate (no.worker), home owner rate (own.rate) and the proportion of African-Americans (afr.pr).

To characterize the cluster a parallel coordinate plot \citep{inselberg1985plane} of the cluster means was used. Figure \ref{cluster.cen} shows the variables used in the horizontal axis and the variable mean of each cluster on the vertical axis. Group 1 are communities with older people, low education level and biggest home owner rate. Group 2 is formed by communities with large proportion of African-Americans and unequal income distribution. Bradenton and Palm Beach (entire group 3) are communities with older people, not working and not African-American. Group 4 is formed by four communities: Charlotte, Grand Forks, Long Beach and State College, all relatively younger people, with high education levels and non equitable income distribution.

A summary of our findings from this graphical exploration of \cite{acs2011} data:
\begin{itemize}
\item Some communities show particular age composition. Bradenton and Palm Beach have a big proportion of elderly people. State College and Grand Forks have a large proportion of young people. The proportion of males shows a similar pattern to females.
\item Education, income distribution and wealth are positively related.
\item Boulder is a well-educated community. Long Beach has a lot of high school dropouts.
\end{itemize}

\section{Exploring attachment \label{explo}}
In this section, we explore the Knight Foundation data. The goal is to look for possible relationships among attachment with different aspects of this data, based on the original variables in the survey instead of the \emph{constructed variables} reported in \cite{full2010} (e.g., openness).

The variable \textbf{CCE} represents the attachment of each person to the place they reside. It is an index between 1 and 5 and it can be decomposed into two components, loyalty and passion. There is also a discrete version of attachment, \textbf{CCEGRP}, a categorical variable with 3 levels: \emph{Attached}, \emph{Neutral} and \emph{Not Attached}. Most of the analysis is done working with CCEGRP (see Figure \ref{pp1} for instance), however in some occasions the CCE is used.

\subsection{Attachment by community}

\begin{figure}[hbpt]
\begin{knitrout}
\definecolor{shadecolor}{rgb}{0.969, 0.969, 0.969}\color{fgcolor}

{\centering \includegraphics[width=\maxwidth]{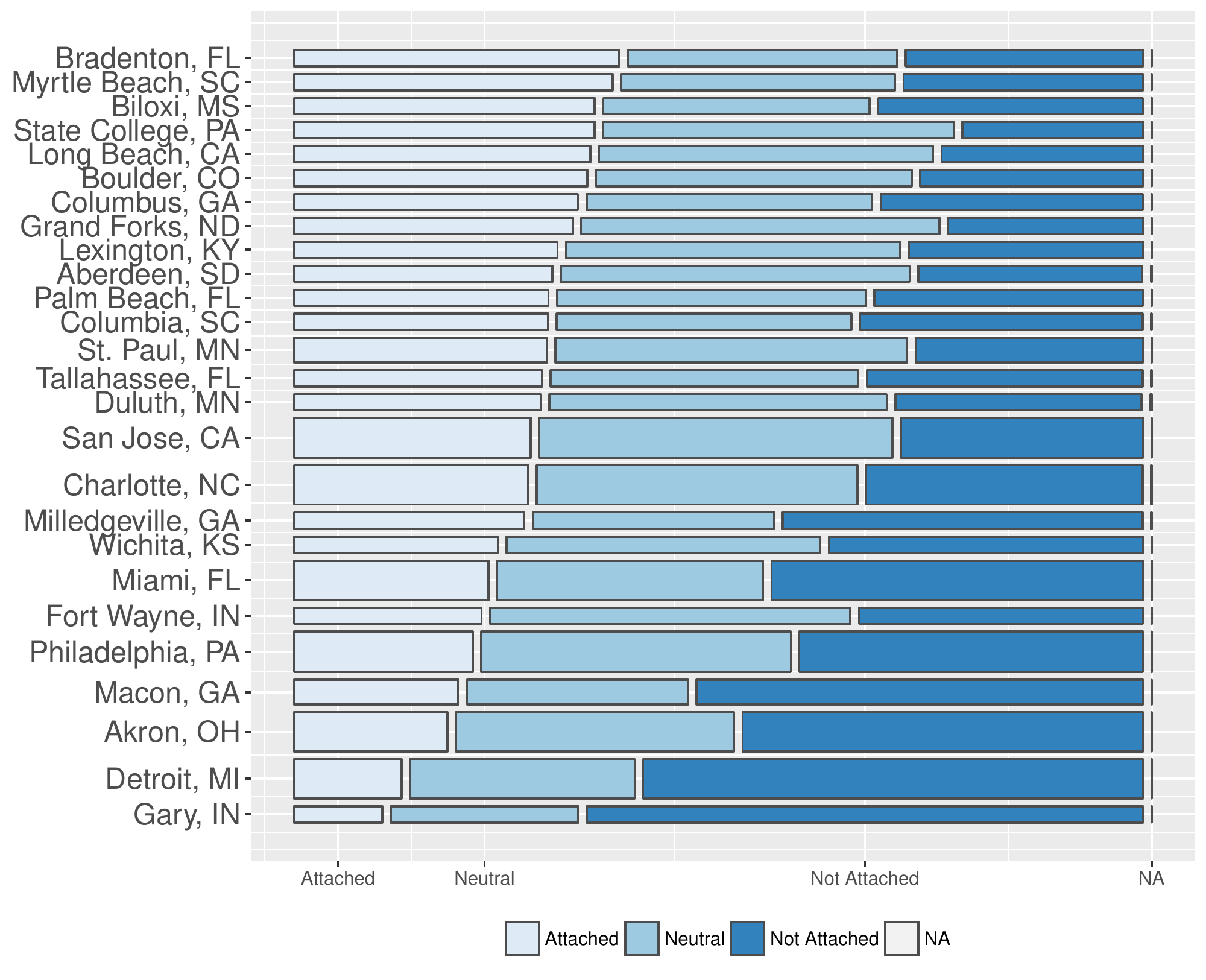} 

}

\end{knitrout}
\caption{Attachment proportion in all 26 communities, sorted from top to bottom by proportion of attached people. In all cases the proportion of attached people is less than 50\%, small communities are more attached than larger ones. \label{comu}}
\end{figure}

The analysis starts by looking at the attachment distribution across all 26 communities. Figure \ref{comu} shows this distribution using a product plot. The width of the box represents the proportion of attachment level within each community and height of the box is related to the size of each community.

Proportion of attached people varies from 10\% to around 40\%, the top three communities in terms of attachment are Bradenton, Myrtle Beach, and Biloxi while the bottom three are Gary, Detroit, and Akron. It is interesting to note that even for the communities with high attached proportions there is still much room to increase level of attachment.

It can be observed that most of the small communities have larger indices of attachment than most of the large communities. In the top fifteen communities the only large community is St Paul. Which could imply community size is important to explain people's  attachment.

\begin{figure}[hbpt]
\begin{knitrout}
\definecolor{shadecolor}{rgb}{0.969, 0.969, 0.969}\color{fgcolor}

{\centering \includegraphics[width=\maxwidth]{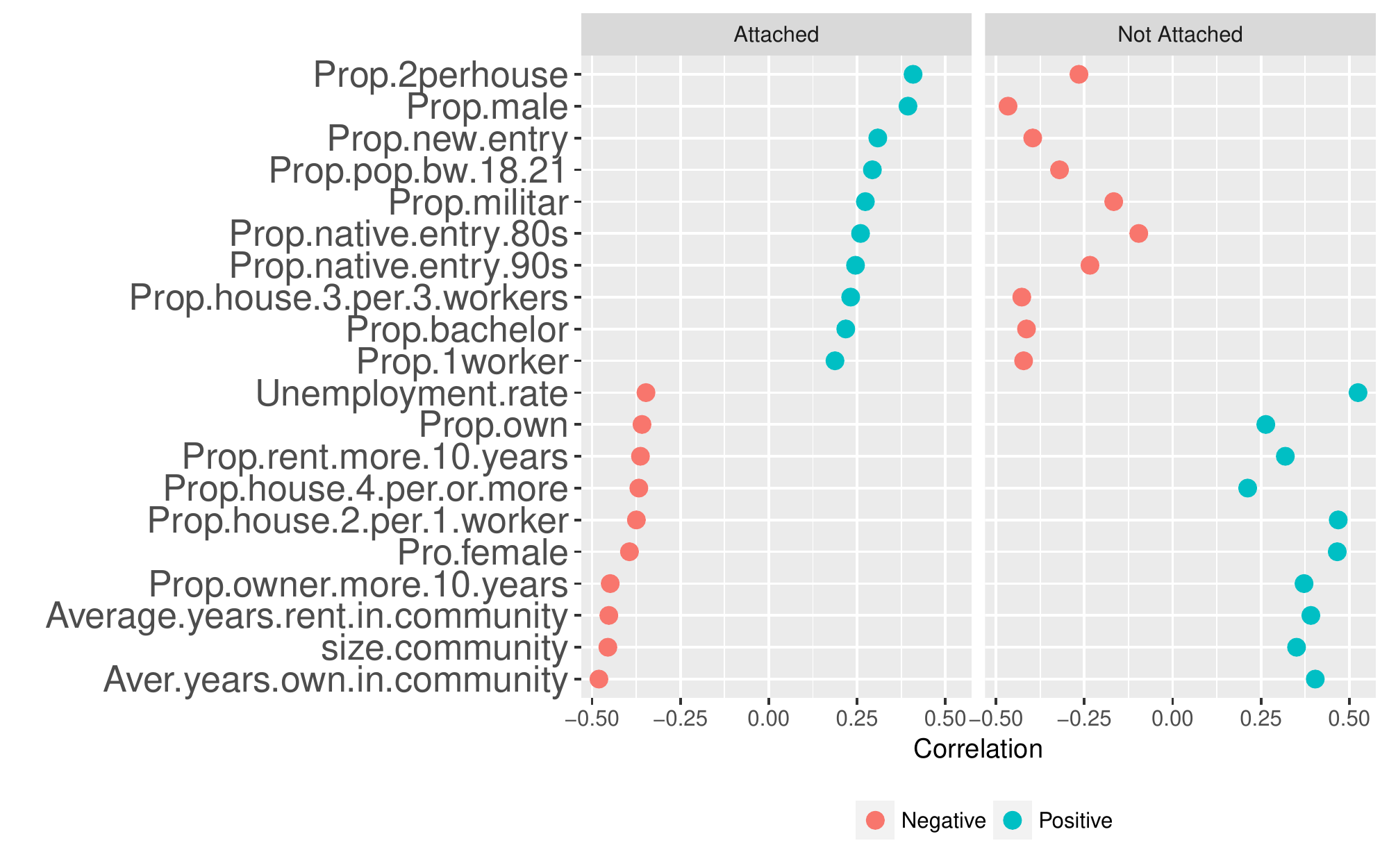} 

}

\end{knitrout}
\caption{Correlations with attached and not attached proportion in order from top to bottom based on correlation with attached\label{corplot}. This plot shows the 10 more relevant variables with positive correlation and 10 with negative correlation with attachment.}
\end{figure}

Figure \ref{corplot} shows the correlation among every variable and the proportion of people attached and not attached per community. It is constructed combining the attachment index in the Knight survey data with the community descriptive variables from Census Bureau. At the community level, proportion of houses with two people, proportion of males, and proportion of people moved into a community within last five years are positively associated with attached, showing correlations between .25 and .35. Interestingly, negative correlations are larger in absolute value, the average years living in the same place (renting or owning) and size of the community are the most negatively correlated variables with attached.

\subsection{Cliques do not click}

In this section our main interest is to characterize people that are attached to their community and see how they are different from not attached people.

Using the original variables (with less than 25\% of missing values), k-means method are used to obtain the clusters of individuals. The variables were converted into a continuous scale. The exploratory search with k-means suggests that after four clusters the within variance reduction is small, so four clusters were used.

\begin{figure}[hbpt]
\begin{knitrout}
\definecolor{shadecolor}{rgb}{0.969, 0.969, 0.969}\color{fgcolor}

{\centering \includegraphics[width=\maxwidth]{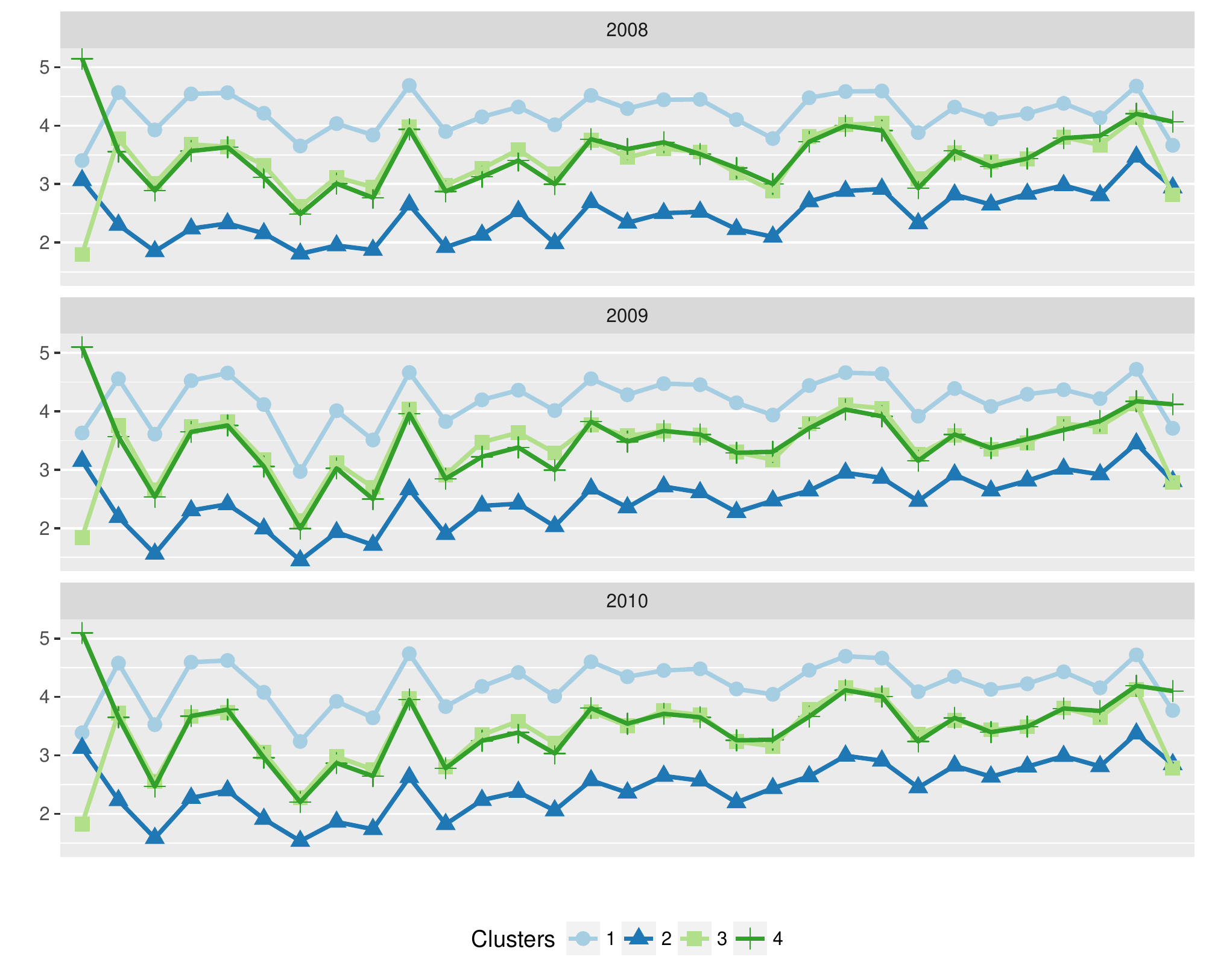} 

}

\end{knitrout}
\caption{Cluster mean points. Mean values for all the variables used to create the clusters in every year. \label{para}}
\end{figure}

Figure \ref{para} is a parallel coordinate plot of the cluster means. The first to notice is that the cluster means are very similar among years.

Cluster 1 shows the biggest mean in most of the variables. These people love their community, and feel it is a good place for different social groups. The only two variables in which this group does not present the largest means are the proportion of close friends and the proportion of family that live in the same community. On the other hand, cluster 2 seems to be the opposite to cluster 1, people in this group are not satisfied with the community where they live.

The only differences among means for clusters 3 and 4 are family and friends indices, i.e. the clique variables. Cluster 4 is characterized by the larger family and friends indices. Clusters 3 and 4 have quite similar mean values for the rest of the variables used to create the clusters.

\begin{figure}[hbpt]
\begin{knitrout}
\definecolor{shadecolor}{rgb}{0.969, 0.969, 0.969}\color{fgcolor}

{\centering \includegraphics[width=\maxwidth]{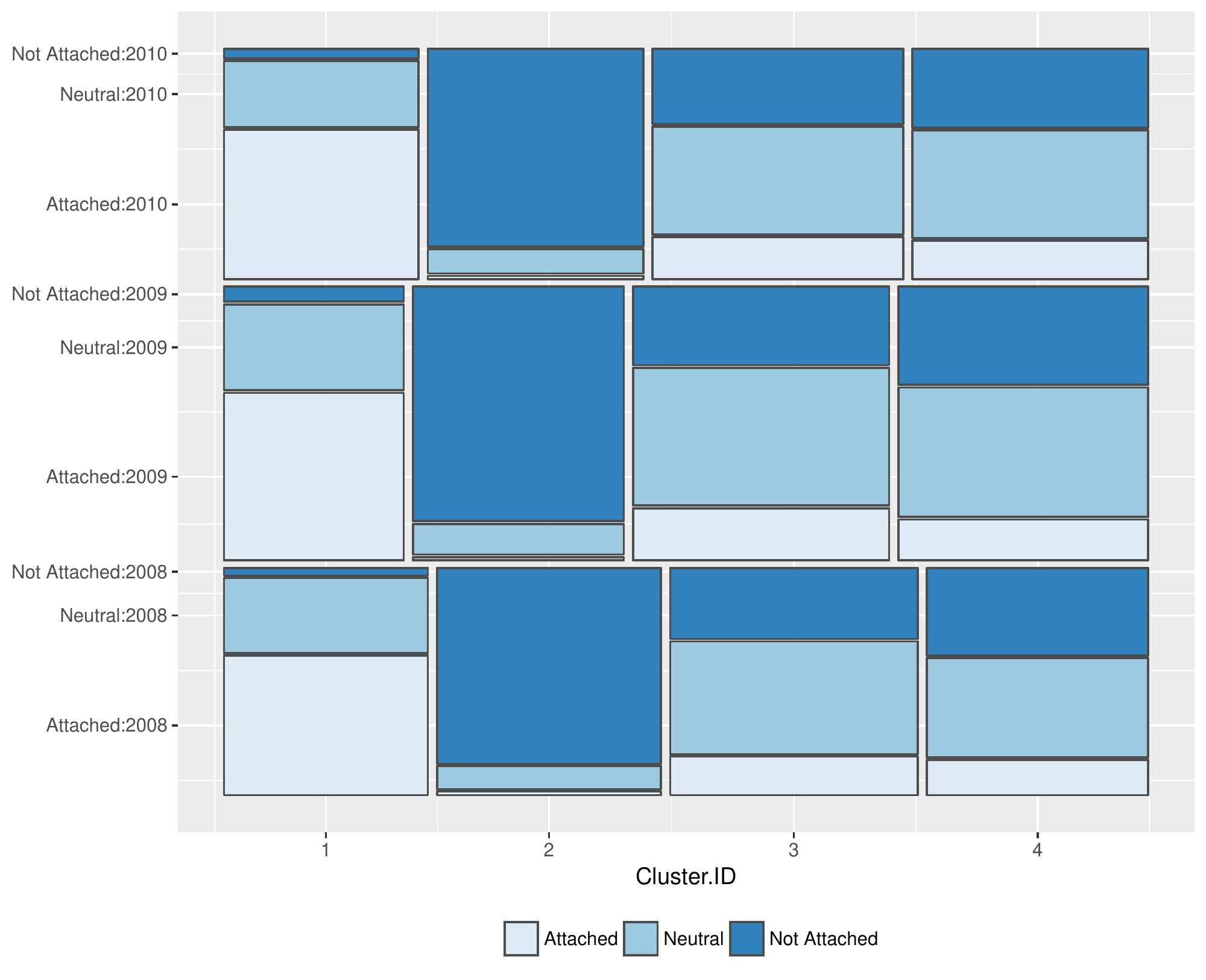} 

}

\end{knitrout}
\caption{Relation between community attachment and k-means cluster solution.\label{clus1}}
\end{figure}

In Figure \ref{clus1} we can see how the clusters are related with the attached group. This plot is another product plot, but now there are three categorical variables combined: year, cluster id and attachment. The area of each box represents the joint proportion of people in one attachment level, cluster id and year. The vertical margin shows that each year is about a third of the total observations. Next within a year the width of the boxes represents the cluster proportions conditional on the year, the four clusters show a pretty uniform distribution. Finally within year and cluster, the height of the box is related to the proportion of each attachment level.  The distribution of attachment is stable across years.

Within cluster 1 the proportion of attached people is more than 75\% every year and there are almost no people who are not attached. On the opposite side, within cluster 2 most of the people are not attached to their community. This is not surprising recalling the cluster means from Figure \ref{para}.
Finally, in clusters 3 and 4 most of the people are in the neutral category. Intuitively having most of your friends and family in the same community where you live will make you feel more attached to it. However, clusters 3 and 4 are similar in terms of attachment but not in terms of family and friends living in the community. It may be that family and friends are not important to attachment.

\subsection{Residence stability}

In this subsection the relationship between attachment and how long people stay in the community is studied. The Knight Foundation questionnaire recorded how many years a person has lived in the current community. A positive relation with attachment is somewhat expected since people who live longer in the community are more likely to be attached to it. However at a community level, it could be the case that attachment level is decreasing if there is no influx of new residents.

\begin{figure}
\begin{knitrout}
\definecolor{shadecolor}{rgb}{0.969, 0.969, 0.969}\color{fgcolor}

{\centering \includegraphics[width=\maxwidth]{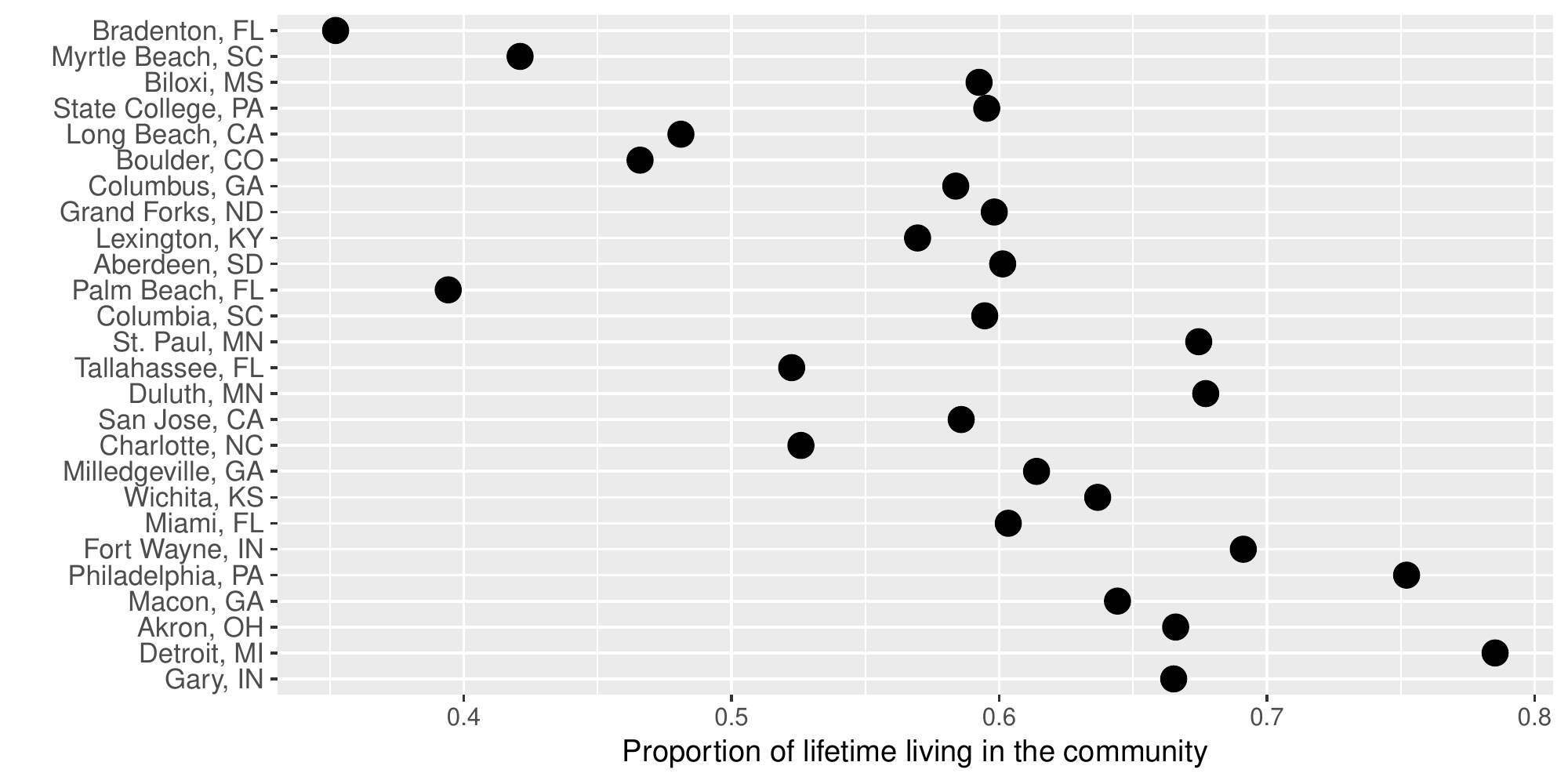} 

}

\end{knitrout}
\caption{Dot chart of communities sorted from top to bottom by attachment against the mean proportion of lifetime spent in the current community (stay). Communities where most of people are living longer are the less attached ones. \label{stay}}
\end{figure}

The proportion of lifetime lived in the community and its mean value for each community was computed. Figure \ref{stay} shows the mean proportion of lifetime lived in the community in a dot chart. As communities are sorted from top to bottom according to attachment proportion, this plot suggests a negative relation among the proportion of attached people and the average proportion of lifetime people have lived there. In other words, communities where new people are moving in regularly are expected to rank higher in terms of attachment than communities where they are not new people.

\begin{table}[ht]
\centering
\caption{Correlation with attachment} 
\label{lifetab}
\begin{tabular}{rllr}
  \hline
 & Data & Description & Correlation \\ 
  \hline
1 & Census Bureau & Average years renting in current home & -0.45 \\ 
  2 & Census Bureau & Average years owning the current home & -0.48 \\ 
  3 & Census Bureau & Prop. of people not borned in the community & 0.13 \\ 
  4 & Knight-Gallup data & Proportion of age in current comunity & -0.53 \\ 
   \hline
\end{tabular}
\end{table}

This relationship is explored in more depth using the attachment index instead of the attach group variable. The correlation between attachment index and all these 3 variables are presented in Table \ref{lifetab}. The correlation between attachment index and average proportion of people living in a community is $ \rho = -0.53$, which is in the same direction suggested by Figure \ref{stay}. From  the \cite{acs2011}  data, the average years for current home owners and renters and the proportion of people who were not born in the community was computed.  The proportion of people attached to a community is negatively correlated with the average years in the same house (as renter or owner) and positively related to the proportion people that are new in the community.

The study of the relationship between attachment and how long people stay in the community might be done also with individual level information.  In the Knight questionnaire people are asked about their ``willingness to move''. By collapsing some of the levels of this variable, it is possible to focus on whether a person wants to move outside its neighborhood or not.

Table \ref{move} shows that among the attached people 19\% would like to move if they had the chance to do it and this proportion increases to 70\% among not attached people. On the other hand, 29 \% of the not attached people would like to stay in the same neighborhood while among attached people this proportion reaches 80\%. This suggests a positive association among the attached group and a longer stay in one community.

\begin{table}[ht]
\centering
\caption{Willingness to move outside your neighborhood} 
\label{move}
\begin{tabular}{rrrr}
  \hline
 & Attached & Neutral & Not Attached \\ 
  \hline
Non Response & 0.12 & 0.26 & 0.16 \\ 
  Move from your neighborhood & 19.21 & 36.47 & 70.46 \\ 
  Stay in your neighborhood & 80.67 & 63.27 & 29.38 \\ 
   \hline
\end{tabular}
\end{table}

\begin{figure}
\begin{knitrout}
\definecolor{shadecolor}{rgb}{0.969, 0.969, 0.969}\color{fgcolor}

{\centering \includegraphics[width=\maxwidth]{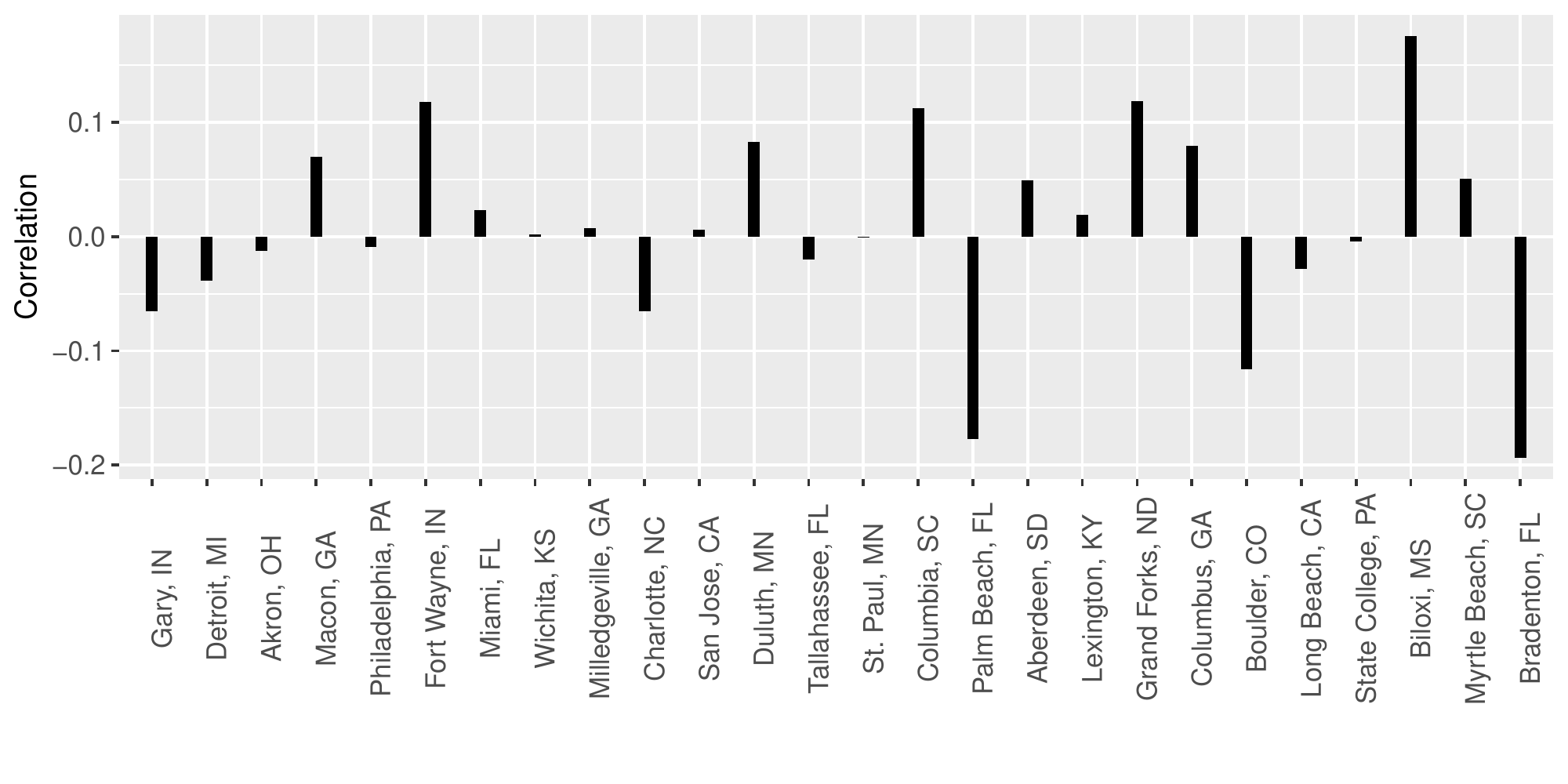} 

}

\end{knitrout}
\caption{ Bar chart of correlations between attachment and lifetime living in the community within each community (ordered by attachment level). Only few communities shows negative correlation between attachment and lifetime living there. \label{wrho}}
\end{figure}

Figure \ref{wrho} shows the correlation among attachment index and proportion of lifetime spent in the community at the individual level for each community. There is not a clear pattern in the correlation and in any case the values are close to the correlation observed at a the community level. None of the  correlations are larger than 0.2 in absolute value and for most communities there is a positive correlation between attached and proportion of lifetime lived in the community.

In summary, the relationship between attachment and how long a person has lived in the community seems to be ambiguous. At the community level, Figure \ref{stay} and Table \ref{lifetab} suggest a negative association, perhaps indicating that communities with elevated proportion of attached people are attractive enough for new people to move into that community. However, at the individual level this relationship is weaker. Table \ref{move} and Figure \ref{wrho} suggest attached people do not want to leave their community, which determines a positive association between higher levels of attachment and how long people live in the community.

\subsection{Other relations}
Two more dimensions were explored to describe the attachment variable. First, the impact of the economic recession on the relationship between attachment and economy index is assessed. Secondly,  how the racial diversity of a community could affect the level of attachment.

Both aspects explored in this subsection are far from a conclusive finding, they are merely possible explanations of some aspect of the data.

\textbf{\emph{The effect of economic condition.}}

\begin{figure}[hbpt]
\begin{knitrout}
\definecolor{shadecolor}{rgb}{0.969, 0.969, 0.969}\color{fgcolor}

{\centering \includegraphics[width=\maxwidth]{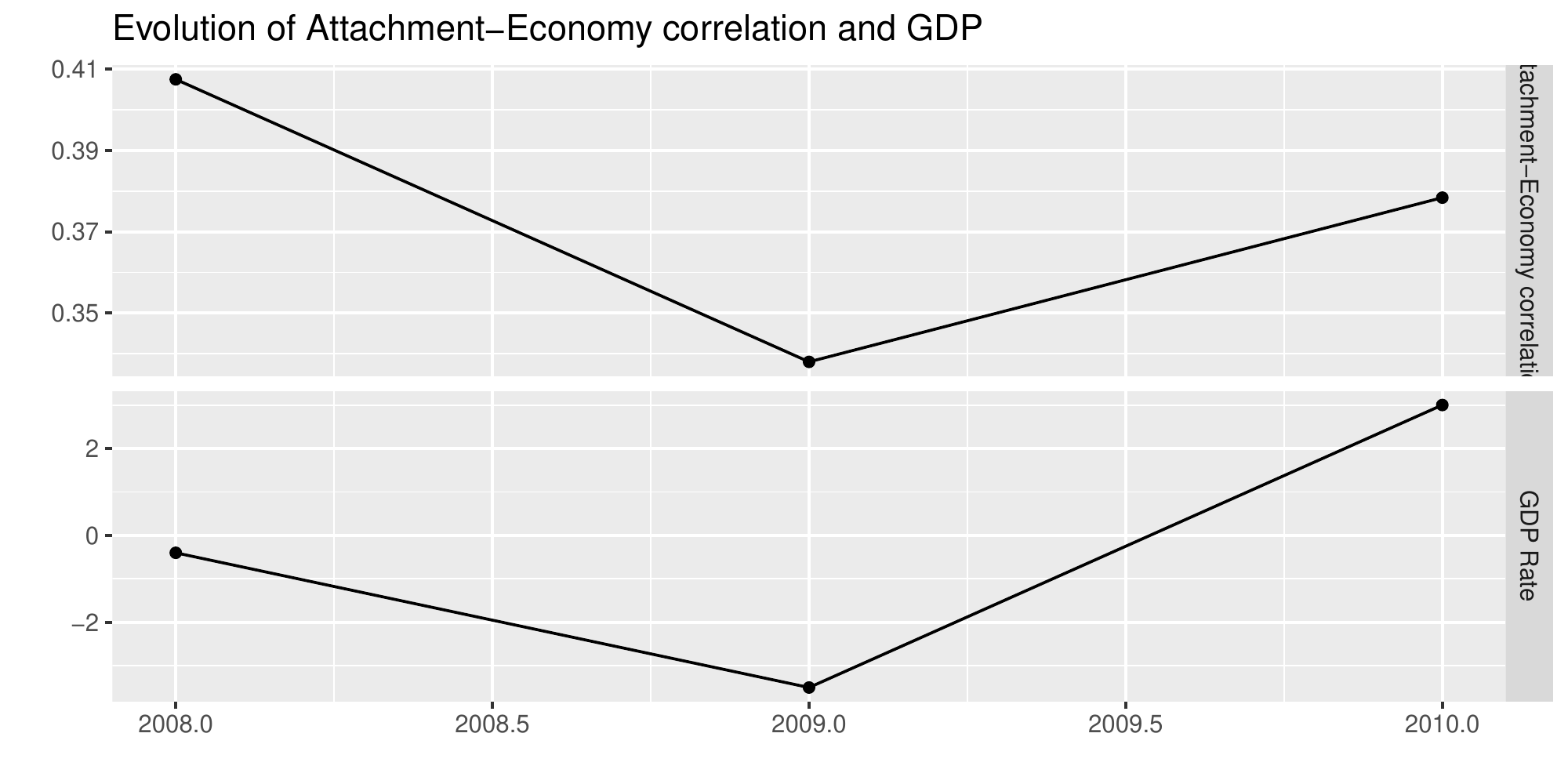} 

}

\end{knitrout}
\caption{Attachment vs economic dimension. Top panel shows the correlation between attachment and economy index against time. Bottom panel shows the GDP growth rate against years. \label{eco}}
\end{figure}

One of the metric variables used by \cite{full2010} is an economy index, which is irrelevant to explain the attachment process in any of the three years of data. Figure \ref{eco} in the top panel shows the correlation between attachment and economic indices for three years of this study. Analysis suggests a relatively low correlation compared to the other metric variables used by \cite{full2010}.

However, the years in which the survey took place were particularly bad years for the U.S. economy. Perhaps the relationship between economy and attachment indices are weakened by the economic recession.

The bottom panel of Figure \ref{eco} presents the U.S. GDP growth rate in 2008, 2009, and 2010. A similar pattern is observed in both panels, suggesting the economic dimension is less related with attachment during the economic crisis.

\textbf{\emph{Race diversity index.}}

Although the largest racial group is white, the 26 surveyed communities have different racial compositions. For instance, the population of Aberdeen is more than 90\% white while San Jose has several racial groups present or Macon where more than 30\% are African-American.

A new variable indicating race was computed as a combination of racial group and ethnicity (Hispanic / non-Hispanic). The resulting variable consists of four categories of racial groups: White, African-American, Hispanic, Other. These are the ethnic/racial groups used in \cite{rushton2008note}, who studied the used of the Blau index as measure of diversity.

The Blau index is defined as $D = 1- \sum p_i^2$ where $p_i$ is the proportion of the $i$-th group in the community.  This measure coincides with the Gini impurity measure used in statistics for decision tree related models\footnote{At the same time, Gini impurity measure is different than the Gini index used for measure income distribution in Section \ref{descri}} (as random forest).
\begin{figure}

\begin{knitrout}
\definecolor{shadecolor}{rgb}{0.969, 0.969, 0.969}\color{fgcolor}

{\centering \includegraphics[width=\maxwidth]{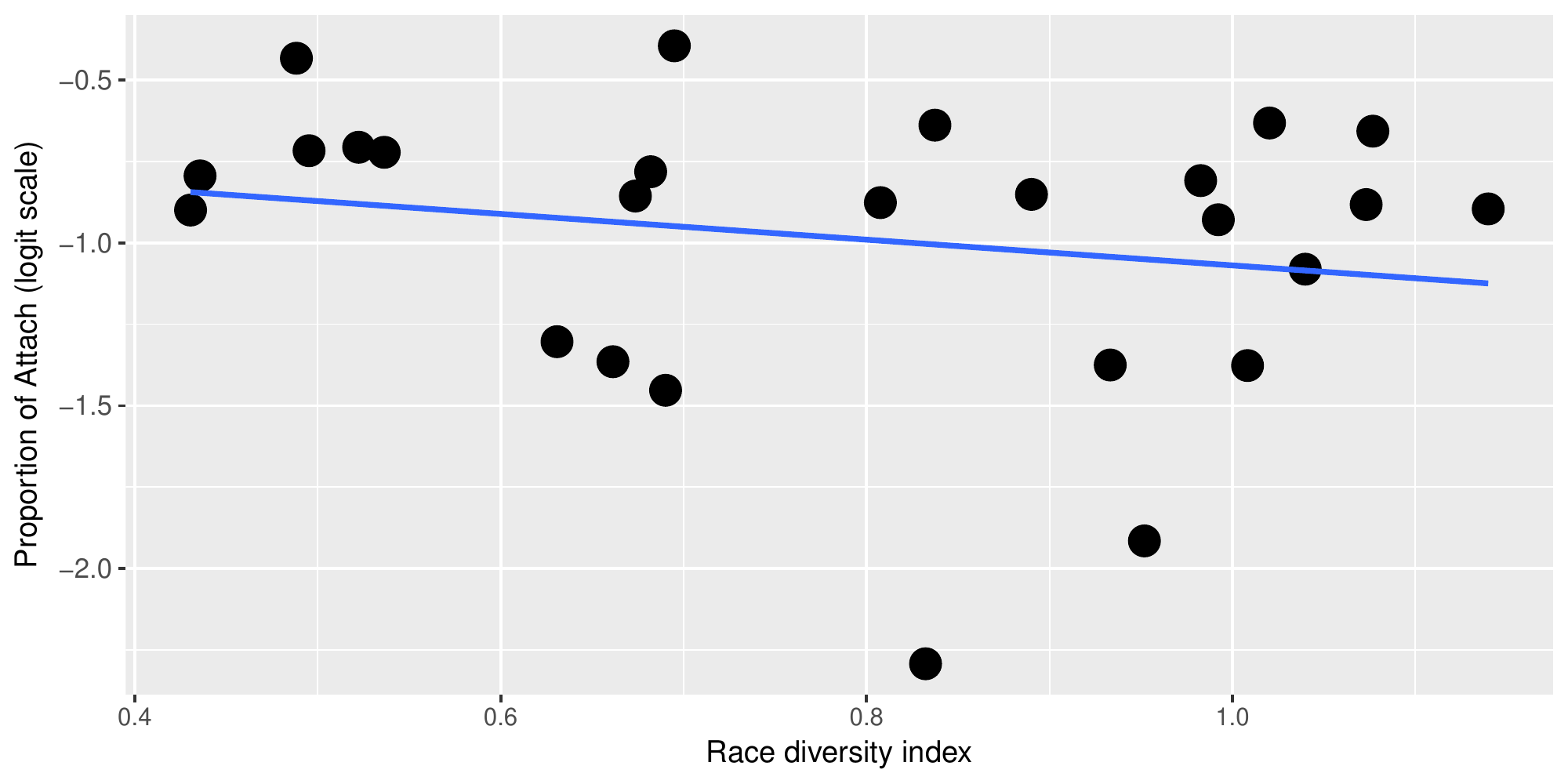} 

}

\end{knitrout}
\caption{Scatter plot of  attachment proportion by race diversity index. \label{raceplot}}
\end{figure}
Figure \ref{raceplot} shows the relation between attached proportion and the race diversity index. The plot suggests a weak negative relationship among attached proportion and racial diversity. It might be possible that for communities where there are different race/ethnic groups present, the attachment to the community is somewhat replaced by the attachment to each particular race group.

\section{Drivers of attachment \label{driver}}

In this section key drivers to explain attachment were found. A random forest classifier was fitted for each community separately. The explanatory variables are all from the original and demographic variables. The attachment group is the response variable.

The most common use for the random forest classifier is for prediction, however the main objective is to identify the key variables to explain the attachment for each community instead of predict the attachment level.

As mentioned in Section \ref{methods}, one output of a random forest classifier is an importance measure for the variables used to construct the forest. The variable importance measures are used to identify the key drivers of the attachment process. Additionally, groups of communities based on these importance measures are created, so communities in the same cluster have a similar structure of key drivers to explain the emotional attachment.

There are two basic measures of variable importance in the \verb randomForest  function within R, only the \emph{mean decrease in Gini} ($DG$) was used. Each time a variable is used within a random forest classifier is associated with a decrease in the impurity of the data nodes, which is a measure with the Gini impurity index. Important variables will achieve larger reduction of impurity, then $DG$ can be used as a variable importance measurement.

\begin{figure}[hbpt]
\begin{knitrout}
\definecolor{shadecolor}{rgb}{0.969, 0.969, 0.969}\color{fgcolor}

{\centering \includegraphics[width=\maxwidth]{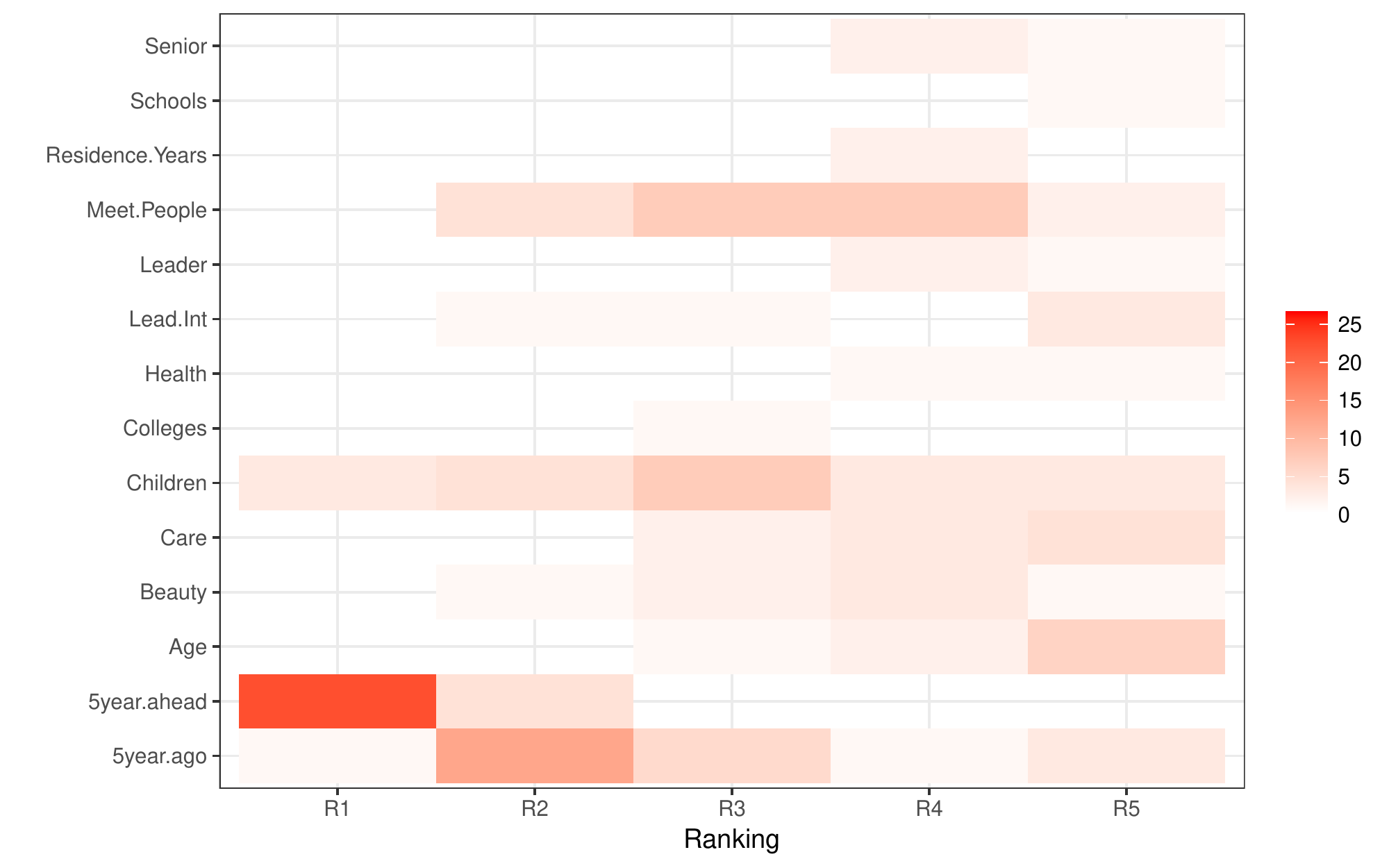} 

}

\end{knitrout}
\caption{Variable importance, color intensity indicates for how many communities a variable appears in each of the first five ranks. For instance, the most important (rank equals to 1) variable is 5.year.ahead for 22 out 26 communities. \label{imp1}}
\end{figure}

Within each community explanatory variables were ranked according to $DG$. Rank 1 represents the variable associated with the largest reduction in the data impurity. Figure \ref{imp1} shows variable importance, and color intensity indicates how many times a variable appears in each of the first five ranks. At a glance it is clear that 5.year.ahead, 5.year.ago and Meet.People, appears as the most important variables in most of the communities. The variable 5.year.ahead indicates the importance of forward looking, i.e. what do you expect from your community in 5 years, and this is the most important driver of attachment.

The variable 5.years.ago represents the perception of the community improvement, when comparing the community today with 5 years ago. Finally, the Meet.People variable indicates if the community is a good place to meet people and make friends.
\begin{figure}[hbpt]
\begin{knitrout}
\definecolor{shadecolor}{rgb}{0.969, 0.969, 0.969}\color{fgcolor}

{\centering \includegraphics[width=\maxwidth]{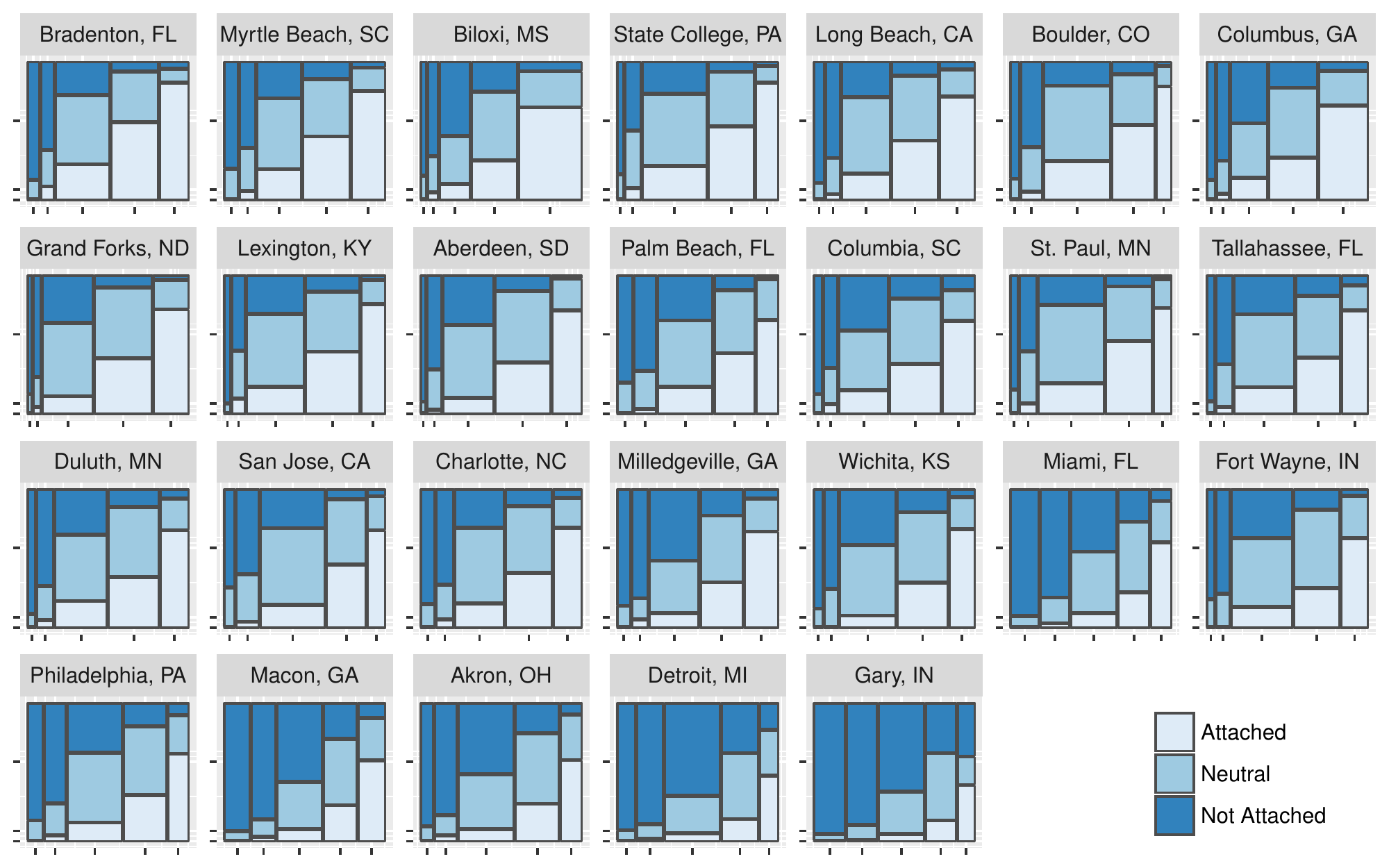} 

}

\end{knitrout}
\caption{Product plot of attachment against forward looking variable (5.year.ahead) by community. \label{fowardplot}}
\end{figure}

What people expect from their community is the most important driver of attachment. Figure \ref{fowardplot} shows the relation of 5.years.ahead variable on each community. Each panel is a community, the width of the boxes represent the proportion of each level of the forward looking variable, and within that box, heights represent the proportion of attachment group. The structure is the same for all communities so this variable is a key driver across all the communities in the study independently of the overall level of attachment.

\begin{figure}
\begin{knitrout}
\definecolor{shadecolor}{rgb}{0.969, 0.969, 0.969}\color{fgcolor}

{\centering \includegraphics[width=\maxwidth]{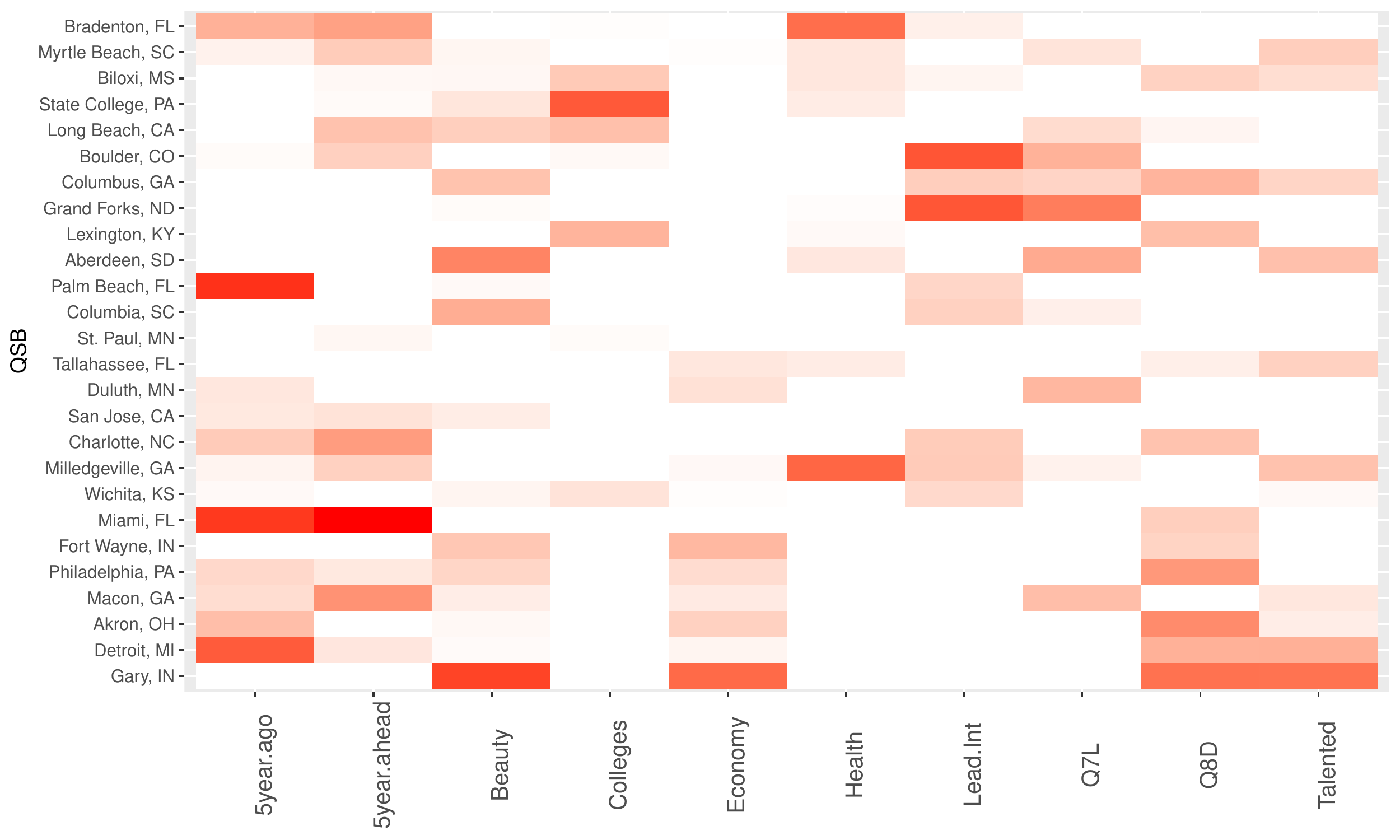} 

}

\end{knitrout}
\caption{Illustration of the differences in important variables by community \label{imp.st}. For each variable, red intensity represent the variable relative importance, i.e. in which community that variable is more important than the average. White indicates variables below importance average.}
\end{figure}

Considering one variable at a time, a different way to look at the variable importance is to explore for which communities that variable is relatively more important. For this, a measure of relative importance was computed. Let $i$ denote the variables and $j$ the communities, then:
\begin{itemize}
  \item First, normalize $DG_{ij}$ as $\tilde{DG_{ij}} = 100\;DG_{ij}/\sum_i DG_{ij}$.
  \item Second, compute relative importance as $rDG_{ij} = \left[ \tilde{DG_{ij}} - \sum_j \tilde{DG_{ij}}/26 \right]_{+} $
\end{itemize}
where $\left[\;\;\right]_{+}$ operation sets the value equal to zero if $\tilde{DG_{ij}}$ is below its mean. In this way, $rDG_{ij}$ will be high if variable $i$ is more important for community $j$ than its mean importance across all communities.

Figure \ref{imp.st} shows the relative importance, red represents $rDG_{ij}$. For each variable (a column in the plot) the red boxes are the communities in which that variable is relatively important. On the vertical axis, communities are sorted from top to bottom by attachment level, so variables where the red colors is concentrated on the top (bottom) is a variable important for the most (least) attached communities.

Some variable can be identified as relatively most important for communities with high levels of attachment compare to communities with lower attachment levels. For instance, the columns corresponding to quality of healthcare (Health) or quality of college and universities (College) in the community show red boxes at the top of the plot.  On the other hand, variables relatively more important for the bottom communities (with lower levels of attachment) are related with economic conditions (economy) and the retrospective vision of the community (5.years.ago). Additionally there are other variables which its relative importance is not associated with attachment level, 5.years.ahead or Beauty presents red boxes for several communities with different levels of attachment.

\begin{figure}
\begin{knitrout}
\definecolor{shadecolor}{rgb}{0.969, 0.969, 0.969}\color{fgcolor}

{\centering \includegraphics[width=\maxwidth]{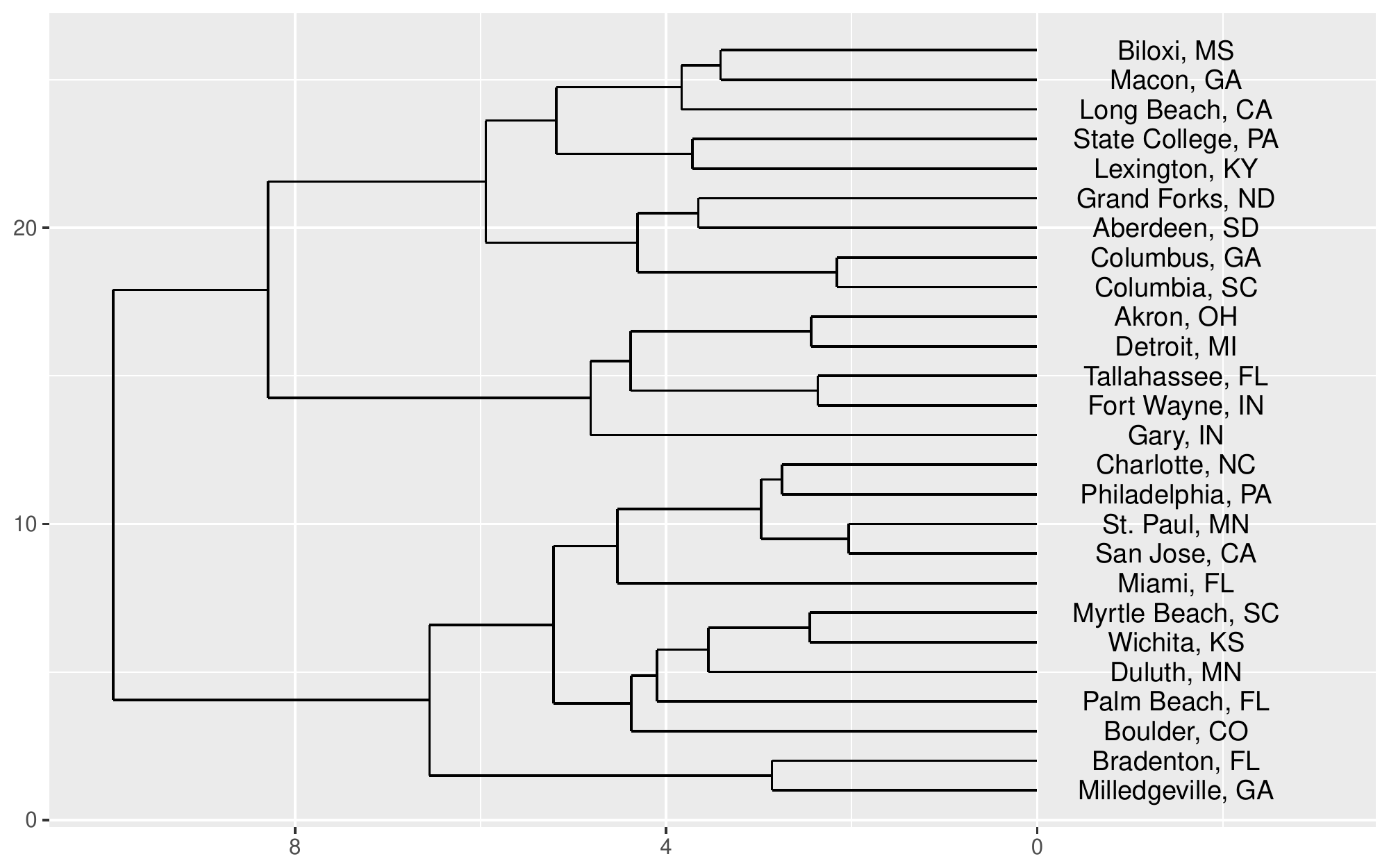} 

}

\end{knitrout}
\caption{Dendrogam of communities base on importance measures. Communities close to each other have similar structure of key drivers of attachment. \label{dendro}}
\end{figure}

Finally it is possible to identify communities with similar structure for its key drivers of attachment. Figure \ref{dendro} shows the dendrogram of communities using the importance measures from the random forest, communities which are grouped together are close in terms if key drivers of attachment. For instance, Detroit, Gary and Akron are the communities with lower levels of attachment and also they are close to each other in terms of the key drivers. However, is possible to find communities with similar structure key drivers but different levels of attachment, like Bradenton and Milledgeville.

\section{Discussion}

In this paper, we showed an exploratory analysis to find interesting facts about the emotional attachment of people to their community. Data from both the Knight Foundation survey and from the Census Bureau were used. Graphical exploration of these data to find patterns that explain the attachment process was done and in addition, statistical analysis for classification and clustering was performed.

The main findings from the analysis are:
\begin{itemize}

\item The 26 communities are a geographically diverse group. Of those, the three main regions are: southeast, northeast and midwest, there are also two communities in California. From the Census Bureau data, we have found that most of the communities show a low proportion of 18-21 year old and also a low proportion of people older than 55 years old. There were just a few communities that have large population of 18-21 year old, (Grand Forks and State College) and larger older population (Palm Beach and Bradenton). There is large variability in education levels and economic conditions across the 26 communities.

\item Attachment proportion differs between communities, but it is always lower than $50\%$. It should be possible for every community to raise the attachment of their people. Using a graphical exploratory data analysis, we have found smaller communities tend to be more attached than larger ones. Also welcoming new residents people is an indicator of a more attached community. This explains that at a community level, there is a negative relationship between attachment and proportion of time lived in the community.

\item Surprisingly, having most of the family or close friends in the same city is not an important driver of the attachment. Individuals that show differences in these dimensions show similar levels of attachment to the community. It is also not important to have people living for a long time in the same community. The correlation between attachment and residence time is close to zero at an individual level. Finally, a preliminary exploration indicates racial diversity does not help to improve attachment to the community.

\item Using random forest model the main key driver for attachment is the future perspective of the community. This variable appears to be important across all 26 communities, especially for those where people are less attached. The better the future perspective the higher the level of attachment.

\item The quality of health and colleges are important on communities with a high proportion of attached people. The economy and past perception in the community are important for unattached communities.
\end{itemize}

\begin{acknowledgements}
Professor Di Cook helped with the analysis and reviewing the paper. Xiaoque Cheng's mergeGui was used to create more amenable initial data. Israel Almodovar, Vianey Leos and Ricardo Martinez help us a lot proofreading our paper for final corrections.
\end{acknowledgements}

\bibliographystyle{spr-chicago}
\bibliography{papersoul}
 
\appendix

\section{Description of the variables used}

\begin{table}[h]
\centering
\caption{Variables used in the paper}
\begin{tabular}{|l|p{10cm}|}\hline
\textbf{Variable} & \textbf{Description} \\ \hline
Afri.pr&  Proportion of african-americans \\
Age & Number of people in age intervals \\
 Age.Fem & Average age among females  \\
Aver.years.rent.in.community&Average of years renting in the community\\
Average.years.own.in.community&Average of years owners in the community\\
 Bachelor Degree(Bachelor) & Proportion of people with Bachelor degree or higher \\
 Beauty & Beauty of physical setting  \\
 Care & How much people in this community care about each other\\
 Children & This community as a place to raise children \\
Colleges & Quality of colleges and universities  \\
Economy & Economic conditions in this community  \\
Education.Level & number of people based on educational attainment, three levels (less high school, high school, some college, bachelor of higher.)  \\
Family & How much of your family lives in your area ? \\
Find.Job & Is now a good time to find a job ?  \\
Friends & How many of your close friends live in your community?   \\
Gay& Gay and lesbian people  \\
GDP& Gross Domestic Product \\
Gender &  gender, male and female  \\
Gini & Gini index to measure income inequality between 0 to 1  \\
Health & Quality of healthcare \\
Housing& Affordable housing \\
Immigrants& Immigrants from other countries\\
Income per capita & Income per-capita in the past 12 months (in 2011 inflation-adjusted dollars in ) \\
Job & Availability of job opportunities  \\
Lead.Int& The leaders in my community represents my interest \\
Leader & The leadership of the elected officials in your city \\
Meet.People & Being a good place to meet people and make friends \\
Minorities& Racial and ethnic minorities \\
New.entry or Prop.new.entry& Proportion of people moved into a community within last 5 years  \\
Night.Life& vibrant nightlife \\
No.workers& Proportion of non working people  \\
Owner.rate & Rate of house owners  \\
Parks & Availability of outdoors, parks\\
Perfect &  This community is the perfect place for people like me  \\
Prop.bachelor&Proportion of people with Bachelor degree or higher \\
Prop.female& Proportion of females\\
Prop.house.2.per.1.worker& Proportion of houses with 2 persons and 1 worker\\
Prop.house.3.per.2.worke&Proportion of houses with 3 persons with 2 workers\\
Prop.house.3.per.3.worker& Proportion of houses with 3 workers\\
Prop.house.3.per.no.workers&Proportion of houses with 3 persons with 3 non workers\\
Prop.house.4.per.or.more&Proportion of house with 4 person or more\\
Prop.male& Proportion of males\\
Prop.militar& Proportion of militars in the community\\
Prop.native.entry.90s& Proportion of US born who enter in the community in 90's \\
Prop.native.entry.prev.80's&Proportion of US born who enter in the community previous to 80's \\
Prop.1worker& Proportion of household with 1 worker\\
Prop.owner.more.10.years& Proportion of owners for more than 10 years\\
Prop.pop.bw.18.21& Proportion of population between 18 and 21 \\
Prop.own& Proportion of owner\\
Prop.rent.more.10.years& Proportion of people renting for more than 10 years\\
Prop.2perhouse& Proportion of houses with two person\\
Prop.unemployed& Proportion of unemployment\\
Prop.white&Proportion of white\\
Proud& I am proud to say I live in this community  \\
Race diversity index & Blau index of race diversity  \\
Refer & How likely are you to recommend this community to a friend or associate as a place to live\\
Residence.Years &  How many years have you lived in this community \\
Reputation &  This community has a good reputation to outsiders or visitors who do not live here \\
Roads & Highway system \\
Safe & On a five-point rating scale, where 5 means completely safe and 1 means not at all safe, how would you rate how safe you feel walking alone at night within a mile of your home \\
 Satisfaction & How satisfied are you with this community as place to live  \\
 Senior & Senior citizens \\
School & Overall quality of public schools \\
size.community& number of people in the community\\
Talented & Young, talented college graduates looking to enter the job market \\
Unemployment.rate& Unemployment rate  \\ \hline
 \end{tabular}
\end{table}

\begin{table}[h]
\centering
\begin{tabular}{|l|p{10cm}|}\hline
1821.Yr& Proportion of people between 18 and 21 years old  \\
5year.ahead & Compare with five years from now. How do you think this community will be as a place to live? \\
5years.ago& How would you compare how this community is as a place to live today, compare to five years ago  \\ \hline
 \end{tabular}
\end{table}

\end{document}